\documentclass[11pt]{article}

\usepackage[
    top=2.5cm,
    bottom=2.5cm,
    left=2.55cm,
    right=2.55cm
]{geometry}
\usepackage{float}

\usepackage{mathtools}
\usepackage{amsmath}
\usepackage{amssymb}
\usepackage{amsfonts}
\usepackage{amstext}
\usepackage{mathrsfs}
\usepackage{placeins}
\usepackage{bbm}
\usepackage{multirow}

\usepackage{graphicx}
\usepackage{tikz}
\usetikzlibrary{calc,math}
\usepackage{multirow}
\tikzset{
	summedLocalHamiltonianAction/.pic = {
		\tikzmath{ \up=0.4;  \step=0.5;   \n=13; \s=0.25;} 
		\coordinate (R) at (0,0);
		\coordinate (L) at ($ (R) - \n*(\step,0)$);
		\draw[rounded corners=3pt,white] ($(L)+(0,\up)$) -- ($(L)+(-\step,\up)$) --  ($(L)+(-\step,-\up)$) -- ($(L)+(0,-\up)$) --cycle;
		\draw[-] ($(L)+ (0,5*\up)$) -- ($(L)+ (0,-5*\up)$) ;
		\foreach \x in {1,...,3}
		\draw[-] ($(L)+ (\x*\step,3*\up)$) -- ($(L)+ (\x*\step,-3*\up)$) ;
		\foreach \x in {6,...,7}
		\draw[-] ($(L)+ (\x*\step,3*\up)$) -- ($(L)+ (\x*\step,-3*\up)$) ;
		\foreach \x in {10,...,12}
		\draw[-] ($(L)+ (\x*\step,3*\up)$) -- ($(L)+ (\x*\step,-3*\up)$) ;
        \draw[line width=0.5mm] ($0.5*(L)+(0,3.5*\up)$) -- ($0.5*(L)+(0,5*\up)$) ;
        \draw[line width=0.5mm] ($0.5*(L)+(0,-3.5*\up)$) -- ($0.5*(L)+(0,-5*\up)$) ;
		\draw[rounded corners=3pt] ($(R)+(0,3*\up)$) -- ($(R)+(\step,3*\up)$) --  ($(R)+(\step,-3*\up)$) -- ($(R)+(0,-3*\up)$) -- cycle ;
		\draw[fill=white,rounded corners=3pt] ($ (L) - (\s,\s) $) rectangle ($ (R) + (\s,\s) $) ;
        \node at ($(L)+(4.5*\step,-2*\up)$) {\Large\dots};
        \node at ($(L)+(4.5*\step,2*\up)$) {\Large\dots};
        \node at ($(L)+(8.5*\step,-2*\up)$) {\Large\dots};
        \node at ($(L)+(8.5*\step,2*\up)$) {\Large\dots};
        \node[scale=4] at ($(L)+(-3*\step,0)$) {$\sum$};
		\draw[fill=white,rounded corners=3pt] ($0.5*(R)+0.5*(L)+(-\step,2.5*\s)$) rectangle ($0.5*(R)+0.5*(L)+(\step,3*\up-1.5*\s)$) ;
		\draw[fill=white,rounded corners=3pt] ($ (L)+(\step-\s,3*\up) $) rectangle ($ (R) + (-\step+\s,3*\up)+ (0,\s) $) ;
		\draw[fill=white,rounded corners=3pt] ($ (L)+(\step-\s,-3*\up-\s) $) rectangle ($ (R) + (-\step+\s,-3*\up) $) ;
        \node at ($0.5*(R)+0.5*(L)$) {$i, \, i+1$};
	}
}

\tikzset{
	stateAction/.pic = {
		\tikzmath{ \up=0.4;  \step=0.5;   \n=13; \s=0.25;} 
		\coordinate (R) at (0,0);
		\coordinate (L) at ($ (R) - \n*(\step,0)$);
		\draw[rounded corners=3pt,white] ($(L)+(0,\up)$) -- ($(L)+(-\step,\up)$) --  ($(L)+(-\step,-\up)$) -- ($(L)+(0,-\up)$) --cycle;
		\draw[-] ($(L)+ (0,5*\up)$) -- ($(L)+ (0,-5*\up)$) ;
		\foreach \x in {1,...,5}
		\draw[-] ($(L)+ (\x*\step,3*\up)$) -- ($(L)+ (\x*\step,-3*\up)$) ;
		\foreach \x in {8,...,12}
		\draw[-] ($(L)+ (\x*\step,3*\up)$) -- ($(L)+ (\x*\step,-3*\up)$) ;
        \draw[line width=0.5mm] ($0.5*(L)+(0,3.5*\up)$) -- ($0.5*(L)+(0,5*\up)$) ;
        \draw[line width=0.5mm] ($0.5*(L)+(0,-3.5*\up)$) -- ($0.5*(L)+(0,-5*\up)$) ;
		\draw[rounded corners=3pt] ($(R)+(0,3*\up)$) -- ($(R)+(\step,3*\up)$) --  ($(R)+(\step,-3*\up)$) -- ($(R)+(0,-3*\up)$) -- cycle ;
		\draw[fill=white,rounded corners=3pt] ($ (L) - (\s,\s) $) rectangle ($ (R) + (\s,\s) $) ;
        \node at ($0.5*(R)+0.5*(L)+(0,-2*\up)$) {\Large\dots};
        \node at ($0.5*(R)+0.5*(L)+(0,2*\up)$) {\Large\dots};
		\draw[fill=black,rounded corners=3pt] ($ (L)+(\step-\s,3*\up) $) rectangle ($ (R) + (-\step+\s,3*\up)+ (0,\s) $) ;
		\draw[fill=black,rounded corners=3pt] ($ (L)+(\step-\s,-3*\up-\s) $) rectangle ($ (R) + (-\step+\s,-3*\up) $) ;
	}
}

\tikzset{
	stateActionMinus1/.pic = {
		\tikzmath{ \up=0.4;  \step=0.5;   \n=12; \s=0.25;} 
		\coordinate (R) at (0,0);
		\coordinate (L) at ($ (R) - \n*(\step,0)$);
		\draw[rounded corners=3pt,white] ($(L)+(0,\up)$) -- ($(L)+(-\step,\up)$) --  ($(L)+(-\step,-\up)$) -- ($(L)+(0,-\up)$) --cycle;
		\draw[-] ($(L)+ (0,5*\up)$) -- ($(L)+ (0,-5*\up)$) ;
		\foreach \x in {1,...,5}
		\draw[-] ($(L)+ (\x*\step,3*\up)$) -- ($(L)+ (\x*\step,-3*\up)$) ;
		\foreach \x in {7,...,11}
		\draw[-] ($(L)+ (\x*\step,3*\up)$) -- ($(L)+ (\x*\step,-3*\up)$) ;
		\draw[-] ($(L)+ (12*\step,5*\up)$) -- ($(L)+ (12*\step,-5*\up)$) ;
        \draw[line width=0.5mm] ($0.5*(L)+(0,3.5*\up)$) -- ($0.5*(L)+(0,5*\up)$) ;
        \draw[line width=0.5mm] ($0.5*(L)+(0,-3.5*\up)$) -- ($0.5*(L)+(0,-5*\up)$) ;
		\draw[fill=white,rounded corners=3pt] ($ (L) - (\s,\s) $) rectangle ($ (R) + (\s,\s) $) ;
        \node at ($0.5*(R)+0.5*(L)+(0,-2*\up)$) {\Large\dots};
        \node at ($0.5*(R)+0.5*(L)+(0,2*\up)$) {\Large\dots};
		\draw[fill=black,rounded corners=3pt] ($ (L)+(\step-\s,3*\up) $) rectangle ($ (R) + (-\step+\s,3*\up)+ (0,\s) $) ;
		\draw[fill=black,rounded corners=3pt] ($ (L)+(\step-\s,-3*\up-\s) $) rectangle ($ (R) + (-\step+\s,-3*\up) $) ;
	}
}

\tikzset{
	rightEdgeLocalHamiltonianActionMinus1/.pic = {
		\tikzmath{ \up=0.4;  \step=0.5;   \n=12; \s=0.25;} 
		\coordinate (R) at (0,0);
		\coordinate (L) at ($ (R) - \n*(\step,0)$);
		\draw[rounded corners=3pt,white] ($(L)+(0,\up)$) -- ($(L)+(-\step,\up)$) --  ($(L)+(-\step,-\up)$) -- ($(L)+(0,-\up)$) --cycle;
		\draw[-] ($(L)+ (0,5*\up)$) -- ($(L)+ (0,-5*\up)$) ;
		\foreach \x in {1,...,5}
		\draw[-] ($(L)+ (\x*\step,3*\up)$) -- ($(L)+ (\x*\step,-3*\up)$) ;
		\foreach \x in {7,...,11}
		\draw[-] ($(L)+ (\x*\step,3*\up)$) -- ($(L)+ (\x*\step,-3*\up)$) ;
		\draw[-] ($(L)+ (12*\step,5*\up)$) -- ($(L)+ (12*\step,-5*\up)$) ;
        \draw[line width=0.5mm] ($0.5*(L)+(0,3.5*\up)$) -- ($0.5*(L)+(0,5*\up)$) ;
        \draw[line width=0.5mm] ($0.5*(L)+(0,-3.5*\up)$) -- ($0.5*(L)+(0,-5*\up)$) ;
		\draw[fill=white,rounded corners=3pt] ($ (L) - (\s,\s) $) rectangle ($ (R) + (\s,\s) $) ;
		\draw[fill=white,rounded corners=3pt] ($0.5*(R)+0.5*(L)+(-\step+5.5*\step,2.5*\s)$) rectangle ($0.5*(R)+0.5*(L)+(\step+5.5*\step,3*\up-1.5*\s)$) ;
        \node at ($0.5*(R)+0.5*(L)+(0,-2*\up)$) {\Large\dots};
        \node at ($0.5*(R)+0.5*(L)+(0,2*\up)$) {\Large\dots};
		\draw[fill=black,rounded corners=3pt] ($ (L)+(\step-\s,3*\up) $) rectangle ($ (R) + (-\step+\s,3*\up)+ (0,\s) $) ;
		\draw[fill=black,rounded corners=3pt] ($ (L)+(\step-\s,-3*\up-\s) $) rectangle ($ (R) + (-\step+\s,-3*\up) $) ;
	}
}

\usepackage{authblk}

\usepackage{comment}
\usepackage{jheppub}

\DeclareFontShape{OT1}{cmr}{mx}{n}{<->cmr10}{}

\newcommand{\CO}{\mathcal{O}}
\newcommand{\CL}{\mathcal{L}}

\newlength\dlf

\makeatletter
\newcommand\blfootnote[1]{%
  \begingroup
  \renewcommand\thefootnote{}\footnote{#1}%
  \addtocounter{footnote}{-1}%
  \endgroup
}
\makeatother

\begin{document}
\def\tikzRhoLow{
    \begin{tikzpicture}[baseline]
        \draw [rounded corners] (-1,-0.5) rectangle (1,0.5);
        \node at (0,0) {$\rho_{N-1}$};

        \foreach \x in {-0.8,0.8}
        {
            \draw (\x,0.5) -- (\x,0.75);
            \draw (\x,-0.5) -- (\x,-0.75);
        }
        \node at (-0.8,1.00) {\scriptsize $\mu_1$};
        \node at (0.8,1.00) {\scriptsize $\mu_{N-1}$};
        \node at (-0.8,-1.00) {\scriptsize $\nu_1$};
        \node at (0.8,-1.00) {\scriptsize $\nu_{N-1}$};
        \foreach \y in {0.6,-0.65}
        {
            \node at (0,\y) {$\cdots$};
        }
    \end{tikzpicture}
}
\def\tikzRhoTrL{
    \begin{tikzpicture}[baseline]
        \draw [rounded corners] (-1.5,-0.5) rectangle (1,0.5);
        \node at (-0.25,0) {$\rho_{N}$};

        \foreach \x in {-0.8,0.8}
        {
            \draw (\x,0.5) -- (\x,0.75);
            \draw (\x,-0.5) -- (\x,-0.75);
        }
        \draw[rounded corners] (-1.3,0.5) -- (-1.3,0.75) -- (-1.75,0.75) -- (-1.75,-0.75) -- (-1.3,-0.75) -- (-1.3,-0.5);
        \node at (-0.8,1.00) {\scriptsize $\mu_1$};
        \node at (0.8,1.00) {\scriptsize $\mu_{N-1}$};
        \node at (-0.8,-1.00) {\scriptsize $\nu_1$};
        \node at (0.8,-1.00) {\scriptsize $\nu_{N-1}$};
        \foreach \y in {0.6,-0.65}
        {
            \node at (0,\y) {$\cdots$};
        }
    \end{tikzpicture}
}
\def\tikzRhoTrR{
    \begin{tikzpicture}[baseline]
        \draw [rounded corners] (-1,-0.5) rectangle (1.5,0.5);
        \node at (0.25,0) {$\rho_{N}$};

        \foreach \x in {0.8,-0.8}
        {
            \draw (\x,0.5) -- (\x,0.75);
            \draw (\x,-0.5) -- (\x,-0.75);
        }
        \draw[rounded corners] (1.3,0.5) -- (1.3,0.75) -- (1.75,0.75) -- (1.75,-0.75) -- (1.3,-0.75) -- (1.3,-0.5);
        \node at (0.8,1.00) {\scriptsize $\mu_{N-1}$};
        \node at (-0.8,1.00) {\scriptsize $\mu_{1}$};
        \node at (-0.8,-1.00) {\scriptsize $\nu_1$};
        \node at (0.8,-1.00) {\scriptsize $\nu_{N-1}$};
        \foreach \y in {0.6,-0.65}
        {
            \node at (0,\y) {$\cdots$};
        }
    \end{tikzpicture}
}

\def\tikzHOdotRho{
    \begin{tikzpicture}[baseline]
        \draw [rounded corners] (-1,-0.5) rectangle (1,0.5);
        \node at (0,0) {$H \odot \rho_{N-1}$};

        \foreach \x in {-0.8,0.8}
        {
            \draw (\x,0.5) -- (\x,0.75);
            \draw (\x,-0.5) -- (\x,-0.75);
        }
        \node at (-0.8,1.00) {\scriptsize $\mu_1$};
        \node at (0.8,1.00) {\scriptsize $\mu_{N-1}$};
        \node at (-0.8,-1.00) {\scriptsize $\nu_1$};
        \node at (0.8,-1.00) {\scriptsize $\nu_{N-1}$};
        \foreach \y in {0.6,-0.65}
        {
            \node at (0,\y) {$\cdots$};
        }
    \end{tikzpicture}
}
\def\tikzHLocalRhoMid{
    \begin{tikzpicture}[baseline]
        \draw [rounded corners] (-1.5,-0.5) rectangle (1,0.5);
        \node at (-0.25,0) {$\rho_{N-1}$};

        \draw (-0.775,0.75) rectangle (0.125,1.25);
        \node at (-0.325,1.00) {$h$};

        \foreach \x in {-1.3,-0.55,-0.05,0.8}
        {
            \draw (\x,0.5) -- (\x,0.75);
        }
        \foreach \x in {-0.55,-0.05}
        {
            \draw (\x,1.25) -- (\x,1.5);
        }
        \draw (0.8,-0.5) -- (0.8,-0.75);
        \draw (-1.3,-0.5) -- (-1.3,-0.75);
        \node at (-0.55,1.75) {\scriptsize $\mu_i$};
        \node at (-0.05,1.75) {\scriptsize $\mu_{i+1}$};
        \node at (0.8,1.00) {\scriptsize $\mu_{N-1}$};
        \node at (-1.3,1.00) {\scriptsize $\mu_{1}$};
        \node at (-1.3,-1.00) {\scriptsize $\nu_1$};
        \node at (0.8,-1.00) {\scriptsize $\nu_{N-1}$};
        \foreach \y in {0.6}
        {
            \foreach \x in {-1.0,0.5}
            {
                \node at (\x,\y) {$\cdots$};
            }
            
        }
        \foreach \y in {-0.65}
        {
            \foreach \x in {-0.25}
            {
                \node at (\x,\y) {$\cdots$};
            }
            
        }
    \end{tikzpicture}
}
\def\tikzHRho{
    \begin{tikzpicture}[baseline]
        \draw [rounded corners] (-1,-0.5) rectangle (1,0.5);
        \node at (0,0) {$\rho_{N-1}$};

        \draw (-1,0.75) rectangle (1,1.25);
        \node at (0,1.00) {$H_\mathrm{bulk}$};

        \foreach \x in {-0.8,0.8}
        {
            \draw (\x,0.5) -- (\x,0.75);
            \draw (\x,1.25) -- (\x,1.5);
            \draw (\x,-0.5) -- (\x,-0.75);
        }
        \node at (-0.8,1.75) {\scriptsize $\mu_1$};
        \node at (0.8,1.75) {\scriptsize $\mu_{N-1}$};
        \node at (-0.8,-1.00) {\scriptsize $\nu_1$};
        \node at (0.8,-1.00) {\scriptsize $\nu_{N-1}$};
        \foreach \y in {0.6,-0.65,1.35}
        {
            \node at (0,\y) {$\cdots$};
        }
    \end{tikzpicture}
}
\def\tikzHLocalRhoL{
    \begin{tikzpicture}[baseline]
        \draw [rounded corners] (-1.5,-0.5) rectangle (1,0.5);
        \node at (-0.25,0) {$\rho_{N}$};

        \draw (-1.5,0.75) rectangle (-0.6,1.25);
        \node at (-1.05,1.00) {$h$};

        \foreach \x in {-1.3,-0.8,0.8}
        {
            \draw (\x,0.5) -- (\x,0.75);
        }
        \draw[rounded corners] (-1.3,1.25) -- (-1.3,1.5) -- (-1.75,1.5) -- (-1.75,-0.75) -- (-1.3,-0.75) -- (-1.3,-0.5);
        \draw (-0.8,1.25) -- (-0.8,1.5);
        \draw (0.8,-0.5) -- (0.8,-0.75);
        \draw (-0.8,-0.5) -- (-0.8,-0.75);
        \node at (-0.8,1.75) {\scriptsize $\mu_1$};
        \node at (0.8,1.00) {\scriptsize $\mu_{N-1}$};
        \node at (-0.8,-1.00) {\scriptsize $\nu_1$};
        \node at (0.8,-1.00) {\scriptsize $\nu_{N-1}$};
        \foreach \y in {0.6,-0.65}
        {
            \node at (0,\y) {$\cdots$};
        }
    \end{tikzpicture}
}
\def\tikzHLocalRhoR{
    \begin{tikzpicture}[baseline]
        \draw [rounded corners] (-1,-0.5) rectangle (1.5,0.5);
        \node at (0.25,0) {$\rho_{N}$};

        \draw (1.5,0.75) rectangle (0.6,1.25);
        \node at (1.05,1.00) {$h$};

        \foreach \x in {1.3,0.8,-0.8}
        {
            \draw (\x,0.5) -- (\x,0.75);
        }
        \draw[rounded corners] (1.3,1.25) -- (1.3,1.5) -- (1.75,1.5) -- (1.75,-0.75) -- (1.3,-0.75) -- (1.3,-0.5);
        \draw (0.8,1.25) -- (0.8,1.5);
        \draw (-0.8,-0.5) -- (-0.8,-0.75);
        \draw (0.8,-0.5) -- (0.8,-0.75);
        \node at (0.8,1.75) {\scriptsize $\mu_{N-1}$};
        \node at (-0.8,1.00) {\scriptsize $\mu_{1}$};
        \node at (-0.8,-1.00) {\scriptsize $\nu_1$};
        \node at (0.8,-1.00) {\scriptsize $\nu_{N-1}$};
        \foreach \y in {0.6,-0.65}
        {
            \node at (0,\y) {$\cdots$};
        }
    \end{tikzpicture}
}

\def\tikzBTensor{
    \begin{tikzpicture}[baseline=-2ex]
        \draw[red,double] (-0.5,0) -- (0.5,0);
        \draw (0,-0.5) -- (0,0);
        \draw[black,fill=black] (0,0) circle (.5ex);
        \node at (0,-0.75) {\scriptsize $\mu$};
        \node at (-0.75,0) {\scriptsize $I$};
        \node at (0.75,0) {\scriptsize $J$};
    \end{tikzpicture}
}

\def\tikzBString{
    \begin{tikzpicture}[baseline=-2ex]
        \draw[red,double] (-1.5,0) -- (0.5,0);
        \draw (0,-0.5) -- (0,0);
        \draw (-1,-0.5) -- (-1,0);
        \draw[black,fill=black] (0,0) circle (.5ex);
        \draw[black,fill=black] (-1,0) circle (.5ex);
        \node at (0.75,-0.025) {\scriptsize $\cdots$};
        \draw[red,double] (1,0) -- (2,0);
        \draw (1.5,-0.5) -- (1.5,0);
        \draw[black,fill=black] (1.5,0) circle (.5ex);
        \node at (-1,-0.75) {\scriptsize $\mu_1$};
        \node at (0,-0.75) {\scriptsize $\mu_2$};
        \node at (1.5,-0.75) {\scriptsize $\mu_N$};
        \node at (-1.75,0) {\scriptsize $I$};
        \node at (2.25,0) {\scriptsize $J$};
    \end{tikzpicture}
}

\def\tikzRhowithB{
    \begin{tikzpicture}[baseline=-0.5ex]
        \draw [rounded corners] (-1.1,-0.3) rectangle (1.5,0.3);
        \node at (0.2,0) {$\rho_{M}$};


        \foreach \x in {-1.0,-0.4,0.8,1.4}
        {
            \draw (\x,0.3) -- (\x,0.6);
            \draw (\x,-0.3) -- (\x,-0.6);
        }
        \draw[red,double,rounded corners] (-0.6,0.8) -- (-0.6,0.6) -- (-0.1,0.6);
        \draw[red,double,rounded corners] (1.0,0.8) -- (1.0,0.6) -- (0.5,0.6);
        \draw[black,fill=black] (-0.4,0.6) circle (.5ex);
        \draw[black,fill=black] (0.8,0.6) circle (.5ex);
        \draw[red,double,rounded corners] (-0.6,-0.8) -- (-0.6,-0.6) -- (-0.1,-0.6);
        \draw[red,double,rounded corners] (1.0,-0.8) -- (1.0,-0.6) -- (0.5,-0.6);
        \draw[black,fill=black] (-0.4,-0.6) circle (.5ex);
        \draw[black,fill=black] (0.8,-0.6) circle (.5ex);
        \node at (-1.0,0.8) {\scriptsize $\mu_1$};
        \node at (1.4,0.8) {\scriptsize $\mu_{N}$};
        \node at (-1.0,-0.8) {\scriptsize $\nu_1$};
        \node at (1.4,-0.8) {\scriptsize $\nu_{N}$};
        \node at (-0.6,1.00) {\scriptsize $I$};
        \node at (1.0,1.00) {\scriptsize $J$};
        \node at (-0.6,-1.00) {\scriptsize $I'$};
        \node at (1.0,-1.00) {\scriptsize $J'$};
        \foreach \x in {0.2}
        {
            \node at (\x,0.5) {$\cdots$};
            \node at (\x,-0.5) {$\cdots$};
        }
    \end{tikzpicture}
}

\def\tikzRhothreeL{
    \begin{tikzpicture}[baseline=-0.5ex]
        \draw [rounded corners] (-0.5,-0.3) rectangle (0.9,0.3);
        \node at (0.2,0) {$\rho_{3}$};


        \foreach \x in {-0.4,0.2,0.8}
        {
            \draw (\x,0.3) -- (\x,0.6);
            \draw (\x,-0.3) -- (\x,-0.6);
        }
        \draw[red,double,rounded corners] (0,0.8) -- (0,0.6) -- (0.5,0.6);
        \draw[red,double,rounded corners] (1,0.8) -- (1,0.6) -- (0.5,0.6);
        \draw[black,fill=black] (0.2,0.6) circle (.5ex);
        \draw[black,fill=black] (0.8,0.6) circle (.5ex);
        \draw[red,double,rounded corners] (0,-0.8) -- (0,-0.6) -- (0.5,-0.6);
        \draw[red,double,rounded corners] (1,-0.8) -- (1,-0.6) -- (0.5,-0.6);
        \draw[black,fill=black] (0.2,-0.6) circle (.5ex);
        \draw[black,fill=black] (0.8,-0.6) circle (.5ex);
    \end{tikzpicture}
}

\def\tikzRhothreeR{
    \begin{tikzpicture}[baseline=-0.5ex]
        \draw [rounded corners] (-0.5,-0.3) rectangle (0.9,0.3);
        \node at (0.2,0) {$\rho_{3}$};


        \foreach \x in {-0.4,0.2,0.8}
        {
            \draw (\x,0.3) -- (\x,0.6);
            \draw (\x,-0.3) -- (\x,-0.6);
        }
        \draw[red,double,rounded corners] (-0.6,0.8) -- (-0.6,0.6) -- (-0.1,0.6);
        \draw[red,double,rounded corners] (0.4,0.8) -- (0.4,0.6) -- (-0.1,0.6);
        \draw[black,fill=black] (0.2,0.6) circle (.5ex);
        \draw[black,fill=black] (-0.4,0.6) circle (.5ex);
        \draw[red,double,rounded corners] (-0.6,-0.8) -- (-0.6,-0.6) -- (-0.1,-0.6);
        \draw[red,double,rounded corners] (0.4,-0.8) -- (0.4,-0.6) -- (-0.1,-0.6);
        \draw[black,fill=black] (0.2,-0.6) circle (.5ex);
        \draw[black,fill=black] (-0.4,-0.6) circle (.5ex);
    \end{tikzpicture}
}

\def\tikzWL{
    \begin{tikzpicture}[baseline=-0.5ex]
        \draw [rounded corners] (-1.5,-0.3) rectangle (1.5,0.3);
        \node at (0,0) {$\rho_{N}$};


        \foreach \x in {-1.4,-1.0,-0.4,0.8,1.4}
        {
            \draw (\x,0.3) -- (\x,0.6);
            \draw (\x,-0.3) -- (\x,-0.6);
        }
        \draw[red,double,rounded corners] (-0.6,0.8) -- (-0.6,0.6) -- (-0.1,0.6);
        \draw[red,double,rounded corners] (1.0,0.8) -- (1.0,0.6) -- (0.5,0.6);
        \draw[black,fill=black] (-0.4,0.6) circle (.5ex);
        \draw[black,fill=black] (0.8,0.6) circle (.5ex);
        \draw[red,double,rounded corners] (-0.6,-0.8) -- (-0.6,-0.6) -- (-0.1,-0.6);
        \draw[red,double,rounded corners] (1.0,-0.8) -- (1.0,-0.6) -- (0.5,-0.6);
        \draw[black,fill=black] (-0.4,-0.6) circle (.5ex);
        \draw[black,fill=black] (0.8,-0.6) circle (.5ex);
        \node at (-1.4,0.8) {\scriptsize $\mu_1$};
        \node at (-1.0,0.8) {\scriptsize $\mu_2$};
        \node at (1.4,0.8) {\scriptsize $\mu_{N}$};
        \node at (-1.4,-0.8) {\scriptsize $\nu_1$};
        \node at (-1.0,-0.8) {\scriptsize $\nu_2$};
        \node at (1.4,-0.8) {\scriptsize $\nu_{N}$};
        \node at (-0.6,1.00) {\scriptsize $I$};
        \node at (1.0,1.00) {\scriptsize $J$};
        \node at (-0.6,-1.00) {\scriptsize $I'$};
        \node at (1.0,-1.00) {\scriptsize $J'$};
        \foreach \x in {0.2}
        {
            \node at (\x,0.5) {$\cdots$};
            \node at (\x,-0.5) {$\cdots$};
        }
    \end{tikzpicture}
}
\def\tikzWR{
    \begin{tikzpicture}[baseline=-0.5ex]
        \draw [rounded corners] (-1.5,-0.3) rectangle (1.5,0.3);
        \node at (0,0) {$\rho_{N}$};


        \foreach \x in {-1.4,-0.8,0.4,1.0,1.4}
        {
            \draw (\x,0.3) -- (\x,0.6);
            \draw (\x,-0.3) -- (\x,-0.6);
        }
        \draw[red,double,rounded corners] (0.6,0.8) -- (0.6,0.6) -- (0.1,0.6);
        \draw[red,double,rounded corners] (-1.0,0.8) -- (-1.0,0.6) -- (-0.5,0.6);
        \draw[black,fill=black] (0.4,0.6) circle (.5ex);
        \draw[black,fill=black] (-0.8,0.6) circle (.5ex);
        \draw[red,double,rounded corners] (0.6,-0.8) -- (0.6,-0.6) -- (0.1,-0.6);
        \draw[red,double,rounded corners] (-1.0,-0.8) -- (-1.0,-0.6) -- (-0.5,-0.6);
        \draw[black,fill=black] (0.4,-0.6) circle (.5ex);
        \draw[black,fill=black] (-0.8,-0.6) circle (.5ex);
        \node at (1.6,0.8) {\scriptsize $\mu_{N}$};
        \node at (1.0,0.8) {\scriptsize $\mu_{N-1}$};
        \node at (-1.4,0.8) {\scriptsize $\mu_{1}$};
        \node at (1.6,-0.8) {\scriptsize $\nu_{N}$};
        \node at (1.0,-0.8) {\scriptsize $\nu_{N-1}$};
        \node at (-1.4,-0.8) {\scriptsize $\nu_{1}$};
        \node at (0.6,1.00) {\scriptsize $J$};
        \node at (-1.0,1.00) {\scriptsize $I$};
        \node at (0.6,-1.00) {\scriptsize $J'$};
        \node at (-1.0,-1.00) {\scriptsize $I'$};
        \foreach \x in {-0.2}
        {
            \node at (\x,0.5) {$\cdots$};
            \node at (\x,-0.5) {$\cdots$};
        }
    \end{tikzpicture}
}
\def\tikzChiDef{
    \begin{tikzpicture}[baseline=1.0ex]
        \draw [rounded corners] (-1.1,-0.3) rectangle (1.5,0.3);
        \node at (0.2,0) {$\rho_{N}$};

        \draw (-1.1,0.6) rectangle (1.5,0.9);
        \node at (0.2,0.75) {\scriptsize $H_\mathrm{bulk}$};

        \foreach \x in {-1.0,-0.4,0.8,1.4}
        {
            \draw (\x,0.3) -- (\x,0.6);
            \draw (\x,0.9) -- (\x,1.2);
            \draw (\x,-0.3) -- (\x,-0.6);
        }
        \draw[red,double,rounded corners] (-0.6,1.4) -- (-0.6,1.2) -- (-0.1,1.2);
        \draw[red,double,rounded corners] (1.0,1.4) -- (1.0,1.2) -- (0.5,1.2);
        \draw[black,fill=black] (-0.4,1.2) circle (.5ex);
        \draw[black,fill=black] (0.8,1.2) circle (.5ex);
        \draw[red,double,rounded corners] (-0.6,-0.8) -- (-0.6,-0.6) -- (-0.1,-0.6);
        \draw[red,double,rounded corners] (1.0,-0.8) -- (1.0,-0.6) -- (0.5,-0.6);
        \draw[black,fill=black] (-0.4,-0.6) circle (.5ex);
        \draw[black,fill=black] (0.8,-0.6) circle (.5ex);
        \node at (-1.0,1.4) {\scriptsize $\mu_1$};
        \node at (1.4,1.4) {\scriptsize $\mu_{N}$};
        \node at (-1.0,-0.8) {\scriptsize $\nu_1$};
        \node at (1.4,-0.8) {\scriptsize $\nu_{N}$};
        \node at (-0.6,1.6) {\scriptsize $I$};
        \node at (1.0,1.6) {\scriptsize $J$};
        \node at (-0.6,-1.00) {\scriptsize $I'$};
        \node at (1.0,-1.00) {\scriptsize $J'$};
        \foreach \x in {0.2}
        {
            \node at (\x,1.1) {$\cdots$};
            \node at (\x,0.4) {$\cdots$};
            \node at (\x,-0.5) {$\cdots$};
        }
    \end{tikzpicture}
}

\def\tikzWLBL{
    \begin{tikzpicture}[baseline=-0.5ex]
        \draw [rounded corners] (-1.5,-0.3) rectangle (1.1,0.3);
        \node at (0,0) {$\omega_{L,N}$};


        \foreach \x in {-1.4,-0.8,1.0}
        {
            \draw (\x,0.3) -- (\x,0.6);
            \draw (\x,-0.3) -- (\x,-0.6);
        }
        \draw[red,double,rounded corners] (-1.0,0.8) -- (-1.0,0.6) -- (-0.2,0.6) -- (-0.2,0.3);
        \draw[red,double,rounded corners] (-1.0,-0.8) -- (-1.0,-0.6) -- (-0.2,-0.6) -- (-0.2,-0.3);
        \draw[red,double,rounded corners] (0.4,0.3) -- (0.4,0.6) ;
        \draw[red,double,rounded corners] (0.4,-0.3) -- (0.4,-0.6) ;
        \draw[black,fill=black] (-0.8,0.6) circle (.5ex);
        \draw[black,fill=black] (-0.8,-0.6) circle (.5ex);
        \node at (-1.0,1.0) {\scriptsize $I$};
        \node at (0.4,0.8) {\scriptsize $J$};
        \node at (-1.4,0.8) {\scriptsize $\mu_1$};
        \node at (1.1,0.8) {\scriptsize $\mu_{N}$};
        \node at (-1.0,-1.0) {\scriptsize $I'$};
        \node at (0.4,-0.8) {\scriptsize $J'$};
        \node at (-1.4,-0.8) {\scriptsize $\nu_1$};
        \node at (1.1,-0.8) {\scriptsize $\nu_{N}$};
    \end{tikzpicture}
}
\def\tikzWRBR{
    \begin{tikzpicture}[baseline=-0.5ex]
        \draw [rounded corners] (-1.1,-0.3) rectangle (1.5,0.3);
        \node at (0,0) {$\omega_{R,N}$};


        \foreach \x in {-1.0,0.8,1.4}
        {
            \draw (\x,0.3) -- (\x,0.6);
            \draw (\x,-0.3) -- (\x,-0.6);
        }
        \draw[red,double,rounded corners] (1.0,0.8) -- (1.0,0.6) -- (0.2,0.6) -- (0.2,0.3);
        \draw[red,double,rounded corners] (1.0,-0.8) -- (1.0,-0.6) -- (0.2,-0.6) -- (0.2,-0.3);
        \draw[red,double,rounded corners] (-0.4,0.3) -- (-0.4,0.6) ;
        \draw[red,double,rounded corners] (-0.4,-0.3) -- (-0.4,-0.6) ;
        \draw[black,fill=black] (0.8,0.6) circle (.5ex);
        \draw[black,fill=black] (0.8,-0.6) circle (.5ex);
        \node at (1.0,1.0) {\scriptsize $J$};
        \node at (-0.4,0.8) {\scriptsize $I$};
        \node at (1.4,0.8) {\scriptsize $\mu_{N}$};
        \node at (-1.1,0.8) {\scriptsize $\mu_{1}$};
        \node at (1.0,-1.0) {\scriptsize $J'$};
        \node at (-0.4,-0.8) {\scriptsize $I'$};
        \node at (1.4,-0.8) {\scriptsize $\nu_N$};
        \node at (-1.1,-0.8) {\scriptsize $\nu_{1}$};
    \end{tikzpicture}
}
\newcommand{\tikzWC}[1]{
    \begin{tikzpicture}[baseline=-0.5ex]
        \draw [rounded corners] (-1.0,-0.3) rectangle (1.0,0.3);
        \node at (0,0) {$\omega_{C,#1}$};


        \foreach \x in {-.9,.9}
        {
            \draw (\x,0.3) -- (\x,0.6);
            \draw (\x,-0.3) -- (\x,-0.6);
        }
        \draw[red,double,rounded corners] (0.3,0.6) -- (0.3,0.3);
        \draw[red,double,rounded corners] (0.3,-0.6) -- (0.3,-0.3);
        \draw[red,double,rounded corners] (-0.3,0.3) -- (-0.3,0.6) ;
        \draw[red,double,rounded corners] (-0.3,-0.3) -- (-0.3,-0.6) ;
        \node at (0.3,0.8) {\scriptsize $J$};
        \node at (-0.3,0.8) {\scriptsize $I$};
        \node at (0.9,0.8) {\scriptsize $\mu_{#1}$};
        \node at (-0.9,0.8) {\scriptsize $\mu_{1}$};
        \node at (0.3,-0.8) {\scriptsize $J'$};
        \node at (-0.3,-0.8) {\scriptsize $I'$};
        \node at (0.9,-0.8) {\scriptsize $\nu_{#1}$};
        \node at (-0.9,-0.8) {\scriptsize $\nu_{1}$};
    \end{tikzpicture}
}
\def\tikzWNKulldefinition{
    \begin{tikzpicture}[baseline=-0.5ex]
        \draw [rounded corners] (-1.0,-0.3) rectangle (1.0,0.3);
        \node at (0,0) {$\omega_{M}$};


        \foreach \x in {-.9,.9}
        {
            \draw (\x,0.3) -- (\x,0.6);
            \draw (\x,-0.3) -- (\x,-0.6);
        }
        \draw[red,double,rounded corners] (0.3,0.6) -- (0.3,0.3);
        \draw[red,double,rounded corners] (0.3,-0.6) -- (0.3,-0.3);
        \draw[red,double,rounded corners] (-0.3,0.3) -- (-0.3,0.6) ;
        \draw[red,double,rounded corners] (-0.3,-0.3) -- (-0.3,-0.6) ;
        \node at (0.3,0.8) {\scriptsize $J$};
        \node at (-0.3,0.8) {\scriptsize $I$};
        \node at (0.9,0.8) {\scriptsize $\mu_{M}$};
        \node at (-0.9,0.8) {\scriptsize $\mu_{1}$};
        \node at (0.3,-0.8) {\scriptsize $J'$};
        \node at (-0.3,-0.8) {\scriptsize $I'$};
        \node at (0.9,-0.8) {\scriptsize $\nu_{M}$};
        \node at (-0.9,-0.8) {\scriptsize $\nu_{1}$};
    \end{tikzpicture}
}
\def\tikzWLTrL{
    \begin{tikzpicture}[baseline=-0.5ex]
        \draw [rounded corners] (-1.5,-0.3) rectangle (1.1,0.3);
        \node at (0,0) {$\omega_{L,N}$};


        \foreach \x in {-0.8,1.0}
        {
            \draw (\x,0.3) -- (\x,0.6);
            \draw (\x,-0.3) -- (\x,-0.6);
        }
        \draw[red,double,rounded corners] (-0.2,0.6) -- (-0.2,0.3);
        \draw[red,double,rounded corners] (-0.2,-0.6) -- (-0.2,-0.3);
        \draw[red,double,rounded corners] (0.4,0.3) -- (0.4,0.6) ;
        \draw[red,double,rounded corners] (0.4,-0.3) -- (0.4,-0.6) ;
        \draw[rounded corners] (-1.4,0.3) -- (-1.4,0.6) -- (-2.0,0.6) -- (-2.0,-0.6) -- (-1.4,-0.6) -- (-1.4,-0.3) ;
        \node at (-0.2,0.8) {\scriptsize $I$};
        \node at (0.4,0.8) {\scriptsize $J$};
        \node at (-0.8,0.8) {\scriptsize $\mu_1$};
        \node at (1.1,0.8) {\scriptsize $\mu_{N-1}$};
        \node at (-0.2,-0.8) {\scriptsize $I'$};
        \node at (0.4,-0.8) {\scriptsize $J'$};
        \node at (-0.8,-0.8) {\scriptsize $\nu_1$};
        \node at (1.1,-0.8) {\scriptsize $\nu_{N-1}$};
    \end{tikzpicture}
}

\def\tikzWRTrR{
    \begin{tikzpicture}[baseline=-0.5ex]
        \draw [rounded corners] (-1.1,-0.3) rectangle (1.5,0.3);
        \node at (0,0) {$\omega_{R,N}$};


        \foreach \x in {-0.8,1.0}
        {
            \draw (\x,0.3) -- (\x,0.6);
            \draw (\x,-0.3) -- (\x,-0.6);
        }
        \draw[red,double,rounded corners] (-0.2,0.6) -- (-0.2,0.3);
        \draw[red,double,rounded corners] (-0.2,-0.6) -- (-0.2,-0.3);
        \draw[red,double,rounded corners] (0.4,0.3) -- (0.4,0.6) ;
        \draw[red,double,rounded corners] (0.4,-0.3) -- (0.4,-0.6) ;
        \draw[rounded corners] (1.4,0.3) -- (1.4,0.6) -- (2.0,0.6) -- (2.0,-0.6) -- (1.4,-0.6) -- (1.4,-0.3) ;
        \node at (-0.2,0.8) {\scriptsize $I$};
        \node at (0.4,0.8) {\scriptsize $J$};
        \node at (-0.8,0.8) {\scriptsize $\mu_1$};
        \node at (1.1,0.8) {\scriptsize $\mu_{N-1}$};
        \node at (-0.2,-0.8) {\scriptsize $I'$};
        \node at (0.4,-0.8) {\scriptsize $J'$};
        \node at (-0.8,-0.8) {\scriptsize $\nu_1$};
        \node at (1.1,-0.8) {\scriptsize $\nu_{N-1}$};
    \end{tikzpicture}
}

\def\tikzWTrRKull{
    \begin{tikzpicture}[baseline=-0.5ex]
        \draw [rounded corners] (-0.9,-0.3) rectangle (1.1,0.3);
        \node at (0,0) {$\omega_{M+1}$};


        \foreach \x in {-0.8}
        {
            \draw (\x,0.3) -- (\x,0.6);
            \draw (\x,-0.3) -- (\x,-0.6);
        }
        \draw[red,double,rounded corners] (-0.2,0.6) -- (-0.2,0.3);
        \draw[red,double,rounded corners] (-0.2,-0.6) -- (-0.2,-0.3);
        \draw[red,double,rounded corners] (0.4,0.3) -- (0.4,0.6) ;
        \draw[red,double,rounded corners] (0.4,-0.3) -- (0.4,-0.6) ;
        \draw[rounded corners] (1.0,0.3) -- (1.0,0.6) -- (1.6,0.6) -- (1.6,-0.6) -- (1.0,-0.6) -- (1.0,-0.3) ;
    \end{tikzpicture}
}

\def\tikzWBRKull{
    \begin{tikzpicture}[baseline=-0.5ex]
        \draw [rounded corners] (-0.9,-0.3) rectangle (1.1,0.3);
        \node at (0,0) {$\omega_{M}$};


        \foreach \x in {-0.8,1.0}
        {
            \draw (\x,0.3) -- (\x,0.6);
            \draw (\x,-0.3) -- (\x,-0.6);
        }
        \draw[red,double,rounded corners] (-0.2,0.6) -- (-0.2,0.3);
        \draw[red,double,rounded corners] (-0.2,-0.6) -- (-0.2,-0.3);
        \draw[red,double,rounded corners] (0.4,0.3) -- (0.4,0.6) -- (1.2,0.6) -- (1.2,0.8);
        \draw[red,double,rounded corners] (0.4,-0.3) -- (0.4,-0.6) -- (1.2,-0.6) -- (1.2,-0.8);
        \draw[black,fill=black] (1.0,0.6) circle (.5ex);
        \draw[black,fill=black] (1.0,-0.6) circle (.5ex);
    \end{tikzpicture}
}

\def\tikzWTrLKull{
    \begin{tikzpicture}[baseline=-0.5ex]
        \draw [rounded corners] (-1.0,-0.3) rectangle (1.1,0.3);
        \node at (0,0) {$\omega_{M+1}$};


        \foreach \x in {1.0}
        {
            \draw (\x,0.3) -- (\x,0.6);
            \draw (\x,-0.3) -- (\x,-0.6);
        }
        \draw[red,double,rounded corners] (-0.2,0.6) -- (-0.2,0.3);
        \draw[red,double,rounded corners] (-0.2,-0.6) -- (-0.2,-0.3);
        \draw[red,double,rounded corners] (0.4,0.3) -- (0.4,0.6) ;
        \draw[red,double,rounded corners] (0.4,-0.3) -- (0.4,-0.6) ;
        \draw[rounded corners] (-0.8,0.3) -- (-0.8,0.6) -- (-1.4,0.6) -- (-1.4,-0.6) -- (-0.8,-0.6) -- (-0.8,-0.3) ;
    \end{tikzpicture}
}
\def\tikzWBLKull{
    \begin{tikzpicture}[baseline=-0.5ex]
        \draw [rounded corners] (-1.0,-0.3) rectangle (1.1,0.3);
        \node at (0,0) {$\omega_{M}$};


        \foreach \x in {-0.8,1.0}
        {
            \draw (\x,0.3) -- (\x,0.6);
            \draw (\x,-0.3) -- (\x,-0.6);
        }
        \draw[red,double,rounded corners] (-0.2,0.3) -- (-0.2,0.6) -- (-1.0,0.6) -- (-1.0,0.8);
        \draw[red,double,rounded corners] (-0.2,-0.3) -- (-0.2,-0.6) -- (-1.0,-0.6) -- (-1.0,-0.8);
        \draw[red,double,rounded corners] (0.4,0.3) -- (0.4,0.6) ;
        \draw[red,double,rounded corners] (0.4,-0.3) -- (0.4,-0.6) ;
        \draw[black,fill=black] (-0.8,0.6) circle (.5ex);
        \draw[black,fill=black] (-0.8,-0.6) circle (.5ex);
    \end{tikzpicture}
}

\def\tikzWfourTrRKull{
    \begin{tikzpicture}[baseline=-0.5ex]
        \draw [rounded corners] (-0.9,-0.3) rectangle (1.1,0.3);
        \node at (0,0) {$\omega_{4}$};


        \foreach \x in {-0.8}
        {
            \draw (\x,0.3) -- (\x,0.6);
            \draw (\x,-0.3) -- (\x,-0.6);
        }
        \draw[red,double,rounded corners] (-0.2,0.6) -- (-0.2,0.3);
        \draw[red,double,rounded corners] (-0.2,-0.6) -- (-0.2,-0.3);
        \draw[red,double,rounded corners] (0.4,0.3) -- (0.4,0.6) ;
        \draw[red,double,rounded corners] (0.4,-0.3) -- (0.4,-0.6) ;
        \draw[rounded corners] (1.0,0.3) -- (1.0,0.6) -- (1.6,0.6) -- (1.6,-0.6) -- (1.0,-0.6) -- (1.0,-0.3) ;
    \end{tikzpicture}
}

\def\tikzWfourTrLKull{
    \begin{tikzpicture}[baseline=-0.5ex]
        \draw [rounded corners] (-1.0,-0.3) rectangle (1.1,0.3);
        \node at (0,0) {$\omega_4$};


        \foreach \x in {1.0}
        {
            \draw (\x,0.3) -- (\x,0.6);
            \draw (\x,-0.3) -- (\x,-0.6);
        }
        \draw[red,double,rounded corners] (-0.2,0.6) -- (-0.2,0.3);
        \draw[red,double,rounded corners] (-0.2,-0.6) -- (-0.2,-0.3);
        \draw[red,double,rounded corners] (0.4,0.3) -- (0.4,0.6) ;
        \draw[red,double,rounded corners] (0.4,-0.3) -- (0.4,-0.6) ;
        \draw[rounded corners] (-0.8,0.3) -- (-0.8,0.6) -- (-1.4,0.6) -- (-1.4,-0.6) -- (-0.8,-0.6) -- (-0.8,-0.3) ;
    \end{tikzpicture}
}

\def\tikzWLTrR{
    \begin{tikzpicture}[baseline=-0.5ex]
        \draw [rounded corners] (-1.5,-0.3) rectangle (1.1,0.3);
        \node at (0,0) {$\omega_{L,N}$};


        \foreach \x in {-1.4,-0.8}
        {
            \draw (\x,0.3) -- (\x,0.6);
            \draw (\x,-0.3) -- (\x,-0.6);
        }
        \draw[red,double,rounded corners] (-0.2,0.6) -- (-0.2,0.3);
        \draw[red,double,rounded corners] (-0.2,-0.6) -- (-0.2,-0.3);
        \draw[red,double,rounded corners] (0.4,0.3) -- (0.4,0.6) ;
        \draw[red,double,rounded corners] (0.4,-0.3) -- (0.4,-0.6) ;
        \draw[rounded corners] (1.0,0.3) -- (1.0,0.6) -- (1.6,0.6) -- (1.6,-0.6) -- (1.0,-0.6) -- (1.0,-0.3) ;
        \node at (-0.2,0.8) {\scriptsize $I$};
        \node at (0.4,0.8) {\scriptsize $J$};
        \node at (-1.4,0.8) {\scriptsize $\mu_1$};
        \node at (-0.8,0.8) {\scriptsize $\mu_{2}$};
        \node at (-0.2,-0.8) {\scriptsize $I'$};
        \node at (0.4,-0.8) {\scriptsize $J'$};
        \node at (-1.4,-0.8) {\scriptsize $\nu_1$};
        \node at (-0.8,-0.8) {\scriptsize $\nu_{2}$};
    \end{tikzpicture}
}
\def\tikzWLBR{
    \begin{tikzpicture}[baseline=-0.5ex]
        \draw [rounded corners] (-1.5,-0.3) rectangle (1.1,0.3);
        \node at (0,0) {$\omega_{L,N-1}$};


        \foreach \x in {-1.4,-0.8,1.0}
        {
            \draw (\x,0.3) -- (\x,0.6);
            \draw (\x,-0.3) -- (\x,-0.6);
        }
        \draw[red,double,rounded corners] (-0.2,0.6) -- (-0.2,0.3);
        \draw[red,double,rounded corners] (-0.2,-0.6) -- (-0.2,-0.3);
        \draw[red,double,rounded corners] (0.4,0.3) -- (0.4,0.6) -- (1.2,0.6) -- (1.2,0.8);
        \draw[red,double,rounded corners] (0.4,-0.3) -- (0.4,-0.6) -- (1.2,-0.6) -- (1.2,-0.8);
        \draw[black,fill=black] (1.0,0.6) circle (.5ex);
        \draw[black,fill=black] (1.0,-0.6) circle (.5ex);
        \node at (-0.2,0.8) {\scriptsize $I$};
        \node at (1.2,1.0) {\scriptsize $J$};
        \node at (-1.4,0.8) {\scriptsize $\mu_1$};
        \node at (-0.8,0.8) {\scriptsize $\mu_{2}$};
        \node at (-0.2,-0.8) {\scriptsize $I'$};
        \node at (1.2,-1.0) {\scriptsize $J'$};
        \node at (-1.4,-0.8) {\scriptsize $\nu_1$};
        \node at (-0.8,-0.8) {\scriptsize $\nu_{2}$};
    \end{tikzpicture}
}
\def\tikzWRTrL{
    \begin{tikzpicture}[baseline=-0.5ex]
        \draw [rounded corners] (-1.0,-0.3) rectangle (1.7,0.3);
        \node at (0,0) {$\omega_{R,N}$};


        \foreach \x in {1.0,1.6}
        {
            \draw (\x,0.3) -- (\x,0.6);
            \draw (\x,-0.3) -- (\x,-0.6);
        }
        \draw[red,double,rounded corners] (-0.2,0.6) -- (-0.2,0.3);
        \draw[red,double,rounded corners] (-0.2,-0.6) -- (-0.2,-0.3);
        \draw[red,double,rounded corners] (0.4,0.3) -- (0.4,0.6) ;
        \draw[red,double,rounded corners] (0.4,-0.3) -- (0.4,-0.6) ;
        \draw[rounded corners] (-0.8,0.3) -- (-0.8,0.6) -- (-1.4,0.6) -- (-1.4,-0.6) -- (-0.8,-0.6) -- (-0.8,-0.3) ;
        \node at (-0.2,0.8) {\scriptsize $I$};
        \node at (0.4,0.8) {\scriptsize $J$};
        \node at (1.8,0.8) {\scriptsize $\mu_{N-1}$};
        \node at (1.1,0.8) {\scriptsize $\mu_{N-2}$};
        \node at (-0.2,-0.8) {\scriptsize $I'$};
        \node at (0.4,-0.8) {\scriptsize $J'$};
        \node at (1.8,-0.8) {\scriptsize $\nu_{N-1}$};
        \node at (1.1,-0.8) {\scriptsize $\nu_{N-2}$};
    \end{tikzpicture}
}
\def\tikzWRBL{
    \begin{tikzpicture}[baseline=-0.5ex]
        \draw [rounded corners] (-1.0,-0.3) rectangle (1.7,0.3);
        \node at (0,0) {$\omega_{R,N-1}$};


        \foreach \x in {-0.8,1.0,1.6}
        {
            \draw (\x,0.3) -- (\x,0.6);
            \draw (\x,-0.3) -- (\x,-0.6);
        }
        \draw[red,double,rounded corners] (-0.2,0.3) -- (-0.2,0.6) -- (-1.0,0.6) -- (-1.0,0.8);
        \draw[red,double,rounded corners] (-0.2,-0.3) -- (-0.2,-0.6) -- (-1.0,-0.6) -- (-1.0,-0.8);
        \draw[red,double,rounded corners] (0.4,0.3) -- (0.4,0.6) ;
        \draw[red,double,rounded corners] (0.4,-0.3) -- (0.4,-0.6) ;
        \draw[black,fill=black] (-0.8,0.6) circle (.5ex);
        \draw[black,fill=black] (-0.8,-0.6) circle (.5ex);
        \node at (-1.0,1.0) {\scriptsize $I$};
        \node at (0.4,0.8) {\scriptsize $J$};
        \node at (1.8,0.8) {\scriptsize $\mu_{N-1}$};
        \node at (1.1,0.8) {\scriptsize $\mu_{N-2}$};
        \node at (-1.0,-1.0) {\scriptsize $I'$};
        \node at (0.4,-0.8) {\scriptsize $J'$};
        \node at (1.8,-0.8) {\scriptsize $\nu_{N-1}$};
        \node at (1.1,-0.8) {\scriptsize $\nu_{N-2}$};
    \end{tikzpicture}
}

\def\tikzChiTrR{
    \begin{tikzpicture}[baseline=-0.5ex]
        \draw [rounded corners] (-1.0,-0.3) rectangle (1.2,0.3);
        \node at (0,0) {$\chi_{N}$};


        \foreach \x in {-0.8}
        {
            \draw (\x,0.3) -- (\x,0.6);
            \draw (\x,-0.3) -- (\x,-0.6);
        }
        \draw[red,double,rounded corners] (-0.2,0.6) -- (-0.2,0.3);
        \draw[red,double,rounded corners] (-0.2,-0.6) -- (-0.2,-0.3);
        \draw[red,double,rounded corners] (0.4,0.3) -- (0.4,0.6) ;
        \draw[red,double,rounded corners] (0.4,-0.3) -- (0.4,-0.6) ;
        \draw[rounded corners] (1.0,0.3) -- (1.0,0.6) -- (1.6,0.6) -- (1.6,-0.6) -- (1.0,-0.6) -- (1.0,-0.3) ;
        \node at (-0.2,0.8) {\scriptsize $I$};
        \node at (0.4,0.8) {\scriptsize $J$};
        \node at (-0.8,0.8) {\scriptsize $\mu_{1}$};
        \node at (-0.2,-0.8) {\scriptsize $I'$};
        \node at (0.4,-0.8) {\scriptsize $J'$};
        \node at (-0.8,-0.8) {\scriptsize $\nu_{1}$};
    \end{tikzpicture}
}
\def\tikzChiBR{
    \begin{tikzpicture}[baseline=-0.5ex]
        \draw [rounded corners] (-1.0,-0.3) rectangle (1.2,0.3);
        \node at (0,0) {$\chi_{N-1}$};


        \foreach \x in {-0.8,1.0}
        {
            \draw (\x,0.3) -- (\x,0.6);
            \draw (\x,-0.3) -- (\x,-0.6);
        }
        \draw[red,double,rounded corners] (-0.2,0.6) -- (-0.2,0.3);
        \draw[red,double,rounded corners] (-0.2,-0.6) -- (-0.2,-0.3);
        \draw[red,double,rounded corners] (0.4,0.3) -- (0.4,0.6) -- (1.2,0.6) -- (1.2,0.8);
        \draw[red,double,rounded corners] (0.4,-0.3) -- (0.4,-0.6) -- (1.2,-0.6) -- (1.2,-0.8);
        \draw[black,fill=black] (1.0,0.6) circle (.5ex);
        \draw[black,fill=black] (1.0,-0.6) circle (.5ex);
        \node at (-0.2,0.8) {\scriptsize $I$};
        \node at (1.2,1.0) {\scriptsize $J$};
        \node at (-0.8,0.8) {\scriptsize $\mu_{1}$};
        \node at (-0.2,-0.8) {\scriptsize $I'$};
        \node at (1.2,-1.0) {\scriptsize $J'$};
        \node at (-0.8,-0.8) {\scriptsize $\nu_{1}$};
    \end{tikzpicture}
}
\def\tikzWRHBR{
    \begin{tikzpicture}[baseline=-0.5ex]
        \draw [rounded corners] (-1.1,-0.3) rectangle (1.5,0.3);
        \node at (0,0) {$\omega_{R,N}$};

        \draw (0.7,0.6) rectangle (1.5,0.9);
        \node at (1.1,0.75) {\scriptsize $h$};

        \foreach \x in {-1.0,0.8}
        {
            \draw (\x,0.3) -- (\x,0.6);
            \draw (\x,-0.3) -- (\x,-0.6);
        }
        \draw (0.8,0.9) -- (0.8,1.2);
        \draw (1.4,0.3) -- (1.4,0.6);
        \draw[red,double,rounded corners] (1.0,1.4) -- (1.0,1.2) -- (0.2,1.2) -- (0.2,0.3);
        \draw[red,double,rounded corners] (1.0,-0.8) -- (1.0,-0.6) -- (0.2,-0.6) -- (0.2,-0.3);
        \draw[red,double,rounded corners] (-0.4,0.3) -- (-0.4,0.6) ;
        \draw[red,double,rounded corners] (-0.4,-0.3) -- (-0.4,-0.6) ;
        \draw[rounded corners] (1.4,0.9) -- (1.4,1.2) -- (2.0,1.2) -- (2.0,-0.6) -- (1.4,-0.6) -- (1.4,-0.3) ;
        \draw[black,fill=black] (0.8,1.2) circle (.5ex);
        \draw[black,fill=black] (0.8,-0.6) circle (.5ex);
        \node at (1.0,1.6) {\scriptsize $J$};
        \node at (-0.4,0.8) {\scriptsize $I$};
        \node at (-1.1,0.8) {\scriptsize $\mu_{1}$};
        \node at (1.0,-1.0) {\scriptsize $J'$};
        \node at (-0.4,-0.8) {\scriptsize $I'$};
        \node at (-1.1,-0.8) {\scriptsize $\nu_{1}$};
    \end{tikzpicture}
}
\def\tikzChiTrL{
    \begin{tikzpicture}[baseline=-0.5ex]
        \draw [rounded corners] (-1.2,-0.3) rectangle (1.0,0.3);
        \node at (0,0) {$\chi_{N}$};


        \foreach \x in {0.8}
        {
            \draw (\x,0.3) -- (\x,0.6);
            \draw (\x,-0.3) -- (\x,-0.6);
        }
        \draw[red,double,rounded corners] (0.2,0.6) -- (0.2,0.3);
        \draw[red,double,rounded corners] (0.2,-0.6) -- (0.2,-0.3);
        \draw[red,double,rounded corners] (-0.4,0.3) -- (-0.4,0.6) ;
        \draw[red,double,rounded corners] (-0.4,-0.3) -- (-0.4,-0.6) ;
        \draw[rounded corners] (-1.0,0.3) -- (-1.0,0.6) -- (-1.6,0.6) -- (-1.6,-0.6) -- (-1.0,-0.6) -- (-1.0,-0.3) ;
        \node at (0.2,0.8) {\scriptsize $J$};
        \node at (-0.4,0.8) {\scriptsize $I$};
        \node at (0.8,0.8) {\scriptsize $\mu_{N}$};
        \node at (0.2,-0.8) {\scriptsize $J'$};
        \node at (-0.4,-0.8) {\scriptsize $I'$};
        \node at (0.8,-0.8) {\scriptsize $\nu_{N}$};
    \end{tikzpicture}
}
\def\tikzChiBL{
    \begin{tikzpicture}[baseline=-0.5ex]
        \draw [rounded corners] (-1.2,-0.3) rectangle (1.0,0.3);
        \node at (0,0) {$\chi_{N-1}$};


        \foreach \x in {0.8,-1.0}
        {
            \draw (\x,0.3) -- (\x,0.6);
            \draw (\x,-0.3) -- (\x,-0.6);
        }
        \draw[red,double,rounded corners] (0.2,0.6) -- (0.2,0.3);
        \draw[red,double,rounded corners] (0.2,-0.6) -- (0.2,-0.3);
        \draw[red,double,rounded corners] (-0.4,0.3) -- (-0.4,0.6) -- (-1.2,0.6) -- (-1.2,0.8);
        \draw[red,double,rounded corners] (-0.4,-0.3) -- (-0.4,-0.6) -- (-1.2,-0.6) -- (-1.2,-0.8);
        \draw[black,fill=black] (-1.0,0.6) circle (.5ex);
        \draw[black,fill=black] (-1.0,-0.6) circle (.5ex);
        \node at (0.2,0.8) {\scriptsize $J$};
        \node at (-1.2,1.0) {\scriptsize $I$};
        \node at (0.8,0.8) {\scriptsize $\mu_{N}$};
        \node at (0.2,-0.8) {\scriptsize $J'$};
        \node at (-1.2,-1.0) {\scriptsize $I'$};
        \node at (0.8,-0.8) {\scriptsize $\nu_{N}$};
    \end{tikzpicture}
}
\def\tikzWLHBL{
    \begin{tikzpicture}[baseline=-0.5ex]
        \draw [rounded corners] (-1.5,-0.3) rectangle (1.1,0.3);
        \node at (0,0) {$\omega_{L,N}$};

        \draw (-0.7,0.6) rectangle (-1.5,0.9);
        \node at (-1.1,0.75) {\scriptsize $h$};

        \foreach \x in {1.0,-0.8}
        {
            \draw (\x,0.3) -- (\x,0.6);
            \draw (\x,-0.3) -- (\x,-0.6);
        }
        \draw (-0.8,0.9) -- (-0.8,1.2);
        \draw (-1.4,0.3) -- (-1.4,0.6);
        \draw[red,double,rounded corners] (-1.0,1.4) -- (-1.0,1.2) -- (-0.2,1.2) -- (-0.2,0.3);
        \draw[red,double,rounded corners] (-1.0,-0.8) -- (-1.0,-0.6) -- (-0.2,-0.6) -- (-0.2,-0.3);
        \draw[red,double,rounded corners] (0.4,0.3) -- (0.4,0.6) ;
        \draw[red,double,rounded corners] (0.4,-0.3) -- (0.4,-0.6) ;
        \draw[rounded corners] (-1.4,0.9) -- (-1.4,1.2) -- (-2.0,1.2) -- (-2.0,-0.6) -- (-1.4,-0.6) -- (-1.4,-0.3) ;
        \draw[black,fill=black] (-0.8,1.2) circle (.5ex);
        \draw[black,fill=black] (-0.8,-0.6) circle (.5ex);
        \node at (-1.0,1.6) {\scriptsize $I$};
        \node at (0.4,0.8) {\scriptsize $J$};
        \node at (1.1,0.8) {\scriptsize $\mu_{N}$};
        \node at (-1.0,-1.0) {\scriptsize $I'$};
        \node at (0.4,-0.8) {\scriptsize $J'$};
        \node at (1.1,-0.8) {\scriptsize $\nu_{N}$};
    \end{tikzpicture}
}

\newcommand{\tikzChiOrHW}[1]{
    \begin{tikzpicture}[baseline=-0.5ex]
        \draw [rounded corners] (-1.0,-0.3) rectangle (1.2,0.3);
        \node at (0.1,0) {#1};


        \foreach \x in {-0.8,1.0}
        {
            \draw (\x,0.3) -- (\x,0.6);
            \draw (\x,-0.3) -- (\x,-0.6);
        }
        \draw[red,double,rounded corners] (-0.2,0.6) -- (-0.2,0.3);
        \draw[red,double,rounded corners] (-0.2,-0.6) -- (-0.2,-0.3);
        \draw[red,double,rounded corners] (0.4,0.3) -- (0.4,0.6) ;
        \draw[red,double,rounded corners] (0.4,-0.3) -- (0.4,-0.6) ;
        \node at (-0.2,0.8) {\scriptsize $I$};
        \node at (0.4,0.8) {\scriptsize $J$};
        \node at (1.0,0.8) {\scriptsize $\mu_{N-1}$};
        \node at (-0.8,0.8) {\scriptsize $\mu_{1}$};
        \node at (-0.2,-0.8) {\scriptsize $I'$};
        \node at (0.4,-0.8) {\scriptsize $J'$};
        \node at (1.0,-0.8) {\scriptsize $\nu_{N-1}$};
        \node at (-0.8,-0.8) {\scriptsize $\nu_{1}$};
    \end{tikzpicture}
}
\def\tikzHLocalWL{
    \begin{tikzpicture}[baseline]
        \draw [rounded corners] (-1.5,-0.3) rectangle (1,0.3);
        \node at (-0.25,0) {$\omega_{L,N}$};

        \draw (-1.5,0.6) rectangle (-0.6,0.9);
        \node at (-1.05,0.75) {\scriptsize $h$};

        \foreach \x in {-1.3,-0.8,0.8}
        {
            \draw (\x,0.3) -- (\x,0.6);
        }
        \draw[rounded corners] (-1.3,0.9) -- (-1.3,1.2) -- (-1.75,1.2) -- (-1.75,-0.6) -- (-1.3,-0.6) -- (-1.3,-0.3);
        \draw (-0.8,0.9) -- (-0.8,1.2);
        \draw (0.8,-0.3) -- (0.8,-0.6);
        \draw (-0.8,-0.3) -- (-0.8,-0.6);
        \node at (-0.8,1.4) {\scriptsize $\mu_1$};
        \node at (0.8,0.8) {\scriptsize $\mu_{N-1}$};
        \node at (-0.8,-0.8) {\scriptsize $\nu_1$};
        \node at (0.8,-0.8) {\scriptsize $\nu_{N-1}$};
        \node at (-0.3,0.8) {\scriptsize $I$};
        \node at (0.3,0.8) {\scriptsize $J$};
        \node at (-0.3,-0.8) {\scriptsize $I'$};
        \node at (0.3,-0.8) {\scriptsize $J'$};
        \foreach \x in {-0.3,0.3}
        {
            \draw [red,double,rounded corners] (\x,0.3) -- (\x,0.6);
            \draw [red,double,rounded corners] (\x,-0.3) -- (\x,-0.6);
        }
    \end{tikzpicture}
}
\def\tikzHLocalWR{
    \begin{tikzpicture}[baseline]
        \draw [rounded corners] (-1,-0.3) rectangle (1.5,0.3);
        \node at (0.25,0) {$\omega_{R,N}$};

        \draw (1.5,0.6) rectangle (0.6,0.9);
        \node at (1.05,0.75) {\scriptsize $h$};

        \foreach \x in {1.3,0.8,-0.8}
        {
            \draw (\x,0.3) -- (\x,0.6);
        }
        \draw[rounded corners] (1.3,0.9) -- (1.3,1.2) -- (1.75,1.2) -- (1.75,-0.6) -- (1.3,-0.6) -- (1.3,-0.3);
        \draw (0.8,0.9) -- (0.8,1.2);
        \draw (-0.8,-0.3) -- (-0.8,-0.6);
        \draw (0.8,-0.3) -- (0.8,-0.6);
        \node at (0.8,1.4) {\scriptsize $\mu_{N-1}$};
        \node at (-0.8,0.8) {\scriptsize $\mu_{1}$};
        \node at (-0.8,-0.8) {\scriptsize $\nu_1$};
        \node at (0.8,-0.8) {\scriptsize $\nu_{N-1}$};
        \node at (-0.3,0.8) {\scriptsize $I$};
        \node at (0.3,0.8) {\scriptsize $J$};
        \node at (-0.3,-0.8) {\scriptsize $I'$};
        \node at (0.3,-0.8) {\scriptsize $J'$};
        \foreach \x in {-0.3,0.3}
        {
            \draw [red,double,rounded corners] (\x,0.3) -- (\x,0.6);
            \draw [red,double,rounded corners] (\x,-0.3) -- (\x,-0.6);
        }
    \end{tikzpicture}
}

\newcommand{\tikzRho}[1]{
    \begin{tikzpicture}[baseline]
        \draw [rounded corners] (-1,-0.5) rectangle (1,0.5);
        \node at (0,0) {$\rho_{#1}$};


        \foreach \x in {-0.8,0.8}
        {
            \draw (\x,0.5) -- (\x,0.75);
            \draw (\x,-0.5) -- (\x,-0.75);
        }
        \node at (-0.8,1.00) {\scriptsize $\mu_1$};
        \node at (0.8,1.00) {\scriptsize $\mu_{#1}$};
        \node at (-0.8,-1.00) {\scriptsize $\nu_1$};
        \node at (0.8,-1.00) {\scriptsize $\nu_{#1}$};
        \foreach \y in {0.6,-0.65}
        {
            \node at (0,\y) {$\cdots$};
        }
    \end{tikzpicture}
}
\newcommand{\tikzH}[3]{
    \begin{tikzpicture}[baseline]
        \draw (-1,-0.25) rectangle (1,0.25);
        \node at (0,0) {$H$};


        \foreach \x in {-0.8,0.8}
        {
            \draw (\x,0.25) -- (\x,0.5);
            \draw (\x,-0.25) -- (\x,-0.5);
        }
        \node at (-0.8,0.75) {\scriptsize $#2_1$};
        \node at (0.8,0.75) {\scriptsize $#2_{#1}$};
        \node at (-0.8,-0.75) {\scriptsize $#3_1$};
        \node at (0.8,-0.75) {\scriptsize $#3_{#1}$};
        \foreach \y in {0.6,-0.65}
        {
            \node at (0,\y) {$\cdots$};
        }
    \end{tikzpicture}
}

\def\tikzHLocalTensorRhoL{
    \begin{tikzpicture}[baseline]
        \draw [rounded corners] (-1.5,-0.5) rectangle (1,0.5);
        \node at (-0.25,0) {$\rho_{N}$};

        \draw (-2.75,-0.25) rectangle (-2,0.25);
        \node at (-2.375,0) {$h$};
        \node at (-2.2,0.75) {\scriptsize $\sigma_1$};
        \node at (-2.2,-0.75) {\scriptsize $\lambda_1$};
        \draw (-2.2,0.25) -- (-2.2,0.5);
        \draw (-2.2,-0.25) -- (-2.2,-0.5);
        \draw[rounded corners] (-2.55,0.25) -- (-2.55,1) -- (-1.3,1) -- (-1.3,0.5) ;
        \draw[rounded corners] (-2.55,-0.25) -- (-2.55,-1) -- (-1.3,-1) -- (-1.3,-0.5) ;

        \draw (-3,0.75) -- (-3,-0.75); 
        \draw (-3.65,0.75) -- (-3.65,-0.75); 
        \node at (-3.325,0) {$\cdots$}; 

        \node at (-3,1) {\scriptsize $\lambda_{N-1}$};
        \node at (-3.65,1) {\scriptsize $\lambda_2$};
        \node at (-3,-1) {\scriptsize $\sigma_{N-1}$};
        \node at (-3.65,-1) {\scriptsize $\sigma_2$};

        \foreach \x in {-0.8,0.8}
        {
            \draw (\x,0.5) -- (\x,0.75);
            \draw (\x,-0.5) -- (\x,-0.75);
        }
        \node at (-0.8,1.00) {\scriptsize $\mu_1$};
        \node at (0.8,1.00) {\scriptsize $\mu_{N-1}$};
        \node at (-0.8,-1.00) {\scriptsize $\nu_1$};
        \node at (0.8,-1.00) {\scriptsize $\nu_{N-1}$};
        \foreach \y in {0.6,-0.65}
        {
            \node at (0,\y) {$\cdots$};
        }
    \end{tikzpicture}
}

\def\tikzHLocalTensorRhoR{
    \begin{tikzpicture}[baseline]
        \draw [rounded corners] (-1,-0.5) rectangle (1.5,0.5);
        \node at (0.25,0) {$\rho_{N}$};

        \draw (2.75,-0.25) rectangle (2,0.25);
        \node at (2.375,0) {$h$};
        \node at (2.2,0.75) {\scriptsize $\sigma_{N-1}$};
        \node at (2.2,-0.75) {\scriptsize $\lambda_{N-1}$};
        \draw (2.2,0.25) -- (2.2,0.5);
        \draw (2.2,-0.25) -- (2.2,-0.5);
        \draw[rounded corners] (2.55,0.25) -- (2.55,1) -- (1.3,1) -- (1.3,0.5) ;
        \draw[rounded corners] (2.55,-0.25) -- (2.55,-1) -- (1.3,-1) -- (1.3,-0.5) ;

        \draw (-1.5,0.75) -- (-1.5,-0.75); 
        \draw (-2.15,0.75) -- (-2.15,-0.75); 
        \node at (-1.825,0) {$\cdots$}; 

        \node at (-1.5,1) {\scriptsize $\lambda_{N-2}$};
        \node at (-2.15,1) {\scriptsize $\lambda_1$};
        \node at (-1.5,-1) {\scriptsize $\sigma_{N-2}$};
        \node at (-2.15,-1) {\scriptsize $\sigma_1$};

        \foreach \x in {-0.8,0.8}
        {
            \draw (\x,0.5) -- (\x,0.75);
            \draw (\x,-0.5) -- (\x,-0.75);
        }
        \node at (-0.8,1.00) {\scriptsize $\mu_1$};
        \node at (0.8,1.00) {\scriptsize $\mu_{N-1}$};
        \node at (-0.8,-1.00) {\scriptsize $\nu_1$};
        \node at (0.8,-1.00) {\scriptsize $\nu_{N-1}$};
        \foreach \y in {0.6,-0.65}
        {
            \node at (0,\y) {$\cdots$};
        }
    \end{tikzpicture}
}

\def\tikzBHB{
    \begin{tikzpicture}[baseline=-3pt]
        \draw (-1,-0.15) rectangle (1,0.15);
        \node at (0,0) {\scriptsize $H_\mathrm{bulk}$};


        \foreach \x in {-0.9,-0.5,0.5,0.9}
        {
            \draw (\x,0.15) -- (\x,0.45);
            \draw (\x,-0.15) -- (\x,-0.45);
        }
        \draw[red,double,rounded corners] (-0.7,0.65) -- (-0.7,0.45) -- (-0.2,0.45);
        \draw[red,double,rounded corners] (0.7,0.65) -- (0.7,0.45) -- (0.2,0.45);
        \draw[black,fill=black] (-0.5,0.45) circle (.5ex);
        \draw[black,fill=black] (0.5,0.45) circle (.5ex);
        \draw[red,double,rounded corners] (-0.7,-0.65) -- (-0.7,-0.45) -- (-0.2,-0.45);
        \draw[red,double,rounded corners] (0.7,-0.65) -- (0.7,-0.45) -- (0.2,-0.45);
        \draw[black,fill=black] (-0.5,-0.45) circle (.5ex);
        \draw[black,fill=black] (0.5,-0.45) circle (.5ex);
        \node at (-0.7,0.85) {\scriptsize $K$};
        \node at (0.7,0.85) {\scriptsize $L$};
        \node at (-0.7,-0.85) {\scriptsize $K'$};
        \node at (0.7,-0.85) {\scriptsize $L'$};
        \node at (-1.0,0.65) {\scriptsize $\lambda_1$};
        \node at (1.2,0.65) {\scriptsize $\lambda_{N-1}$};
        \node at (-1.0,-0.65) {\scriptsize $\sigma_1$};
        \node at (1.2,-0.65) {\scriptsize $\sigma_{N-1}$};
        \foreach \y in {-0.48,0.42}
        {
            \node at (0,\y) {\scriptsize $\cdots$};
        }
    \end{tikzpicture}
}
\def\tikzHLocalTensorWL{
    \begin{tikzpicture}[baseline=-3pt]
        \draw [rounded corners] (-1.5,-0.3) rectangle (1,0.3);
        \node at (-0.25,0) {$\omega_{L,N}$};

        \draw (-2.55,-0.15) rectangle (-1.8,0.15);
        \node at (-2.175,0) {\scriptsize $h$};
        \node at (-2.0,0.65) {\scriptsize $\sigma_1$};
        \node at (-2.0,-0.65) {\scriptsize $\lambda_1$};
        \node at (-2.85,0.8) {\scriptsize $\sigma_{N-1}$};
        \node at (-2.85,-0.8) {\scriptsize $\lambda_{N-1}$};
        \draw (-2.85,-.6) -- (-2.85,.6);
        \draw (-2.0,0.15) -- (-2.0,0.45);
        \draw (-2.0,-0.15) -- (-2.0,-0.45);
        \draw[rounded corners] (-2.35,0.15) -- (-2.35,1) -- (-1.3,1) -- (-1.3,0.3) ;
        \draw[rounded corners] (-2.35,-0.15) -- (-2.35,-1) -- (-1.3,-1) -- (-1.3,-0.3) ;

        \foreach \x in {-0.8,0.8}
        {
            \draw (\x,0.3) -- (\x,0.6);
            \draw (\x,-0.3) -- (\x,-0.6);
        }
        \node at (-0.8,0.8) {\scriptsize $\mu_1$};
        \node at (0.8,0.8) {\scriptsize $\mu_{N-1}$};
        \node at (-0.8,-0.8) {\scriptsize $\nu_1$};
        \node at (0.8,-0.8) {\scriptsize $\nu_{N-1}$};
        \node at (-0.3,0.8) {\scriptsize $I$};
        \node at (0.3,0.8) {\scriptsize $J$};
        \node at (-0.3,-0.8) {\scriptsize $I'$};
        \node at (0.3,-0.8) {\scriptsize $J'$};
        \foreach \x in {-0.3,0.3}
        {
            \draw [red,double,rounded corners] (\x,0.3) -- (\x,0.6);
            \draw [red,double,rounded corners] (\x,-0.3) -- (\x,-0.6);
        }
    \end{tikzpicture}
}
\def\tikzHLocalTensorWR{
    \begin{tikzpicture}[baseline=-3pt]
        \draw [rounded corners] (-1,-0.3) rectangle (1.5,0.3);
        \node at (0.25,0) {$\omega_{R,N}$};

        \draw (2.55,-0.15) rectangle (1.8,0.15);
        \node at (2.175,0) {\scriptsize $h$};
        \node at (2.0,0.65) {\scriptsize $\sigma_{N-1}$};
        \node at (2.0,-0.65) {\scriptsize $\lambda_{N-1}$};
        \draw (2.0,0.15) -- (2.0,0.45);
        \draw (2.0,-0.15) -- (2.0,-0.45);
        \draw (-1.25,-.6) -- (-1.25,.6);
        \node at (-1.25,0.8) {\scriptsize $\sigma_{1}$};
        \node at (-1.25,-0.8) {\scriptsize $\lambda_{1}$};
        \draw[rounded corners] (2.35,0.15) -- (2.35,1) -- (1.3,1) -- (1.3,0.3) ;
        \draw[rounded corners] (2.35,-0.15) -- (2.35,-1) -- (1.3,-1) -- (1.3,-0.3) ;

        \foreach \x in {-0.8,0.8}
        {
            \draw (\x,0.3) -- (\x,0.6);
            \draw (\x,-0.3) -- (\x,-0.6);
        }
        \node at (-0.8,0.8) {\scriptsize $\mu_1$};
        \node at (0.8,0.8) {\scriptsize $\mu_{N-1}$};
        \node at (-0.8,-0.8) {\scriptsize $\nu_1$};
        \node at (0.8,-0.8) {\scriptsize $\nu_{N-1}$};
        \node at (-0.3,0.8) {\scriptsize $I$};
        \node at (0.3,0.8) {\scriptsize $J$};
        \node at (-0.3,-0.8) {\scriptsize $I'$};
        \node at (0.3,-0.8) {\scriptsize $J'$};
        \foreach \x in {-0.3,0.3}
        {
            \draw [red,double,rounded corners] (\x,0.3) -- (\x,0.6);
            \draw [red,double,rounded corners] (\x,-0.3) -- (\x,-0.6);
        }
    \end{tikzpicture}
}
\def\tikzTrsfMat{
    \begin{tikzpicture}[baseline=-3pt]
        \foreach \x in {-0.9,-0.5,0.5,0.9}
        {
            \draw (\x,0.2) -- (\x,-0.2);
        }
        \draw[red,double,rounded corners] (-1.1,0.4) -- (-1.1,0.2) -- (-0.3,0.2) ;
        \draw[red,double,rounded corners] (1.1,0.4) -- (1.1,0.2) -- (0.3,0.2) ;
        \draw[red,double,rounded corners] (-1.1,-0.4) -- (-1.1,-0.2) -- (-0.3,-0.2) ;
        \draw[red,double,rounded corners] (1.1,-0.4) -- (1.1,-0.2) -- (0.3,-0.2) ;
        \draw[black,fill=black] (-0.9,0.2) circle (.5ex);
        \draw[black,fill=black] (-0.5,0.2) circle (.5ex);
        \draw[black,fill=black] (0.5,0.2) circle (.5ex);
        \draw[black,fill=black] (0.9,0.2) circle (.5ex);
        \draw[black,fill=black] (-0.9,-0.2) circle (.5ex);
        \draw[black,fill=black] (-0.5,-0.2) circle (.5ex);
        \draw[black,fill=black] (0.5,-0.2) circle (.5ex);
        \draw[black,fill=black] (0.9,-0.2) circle (.5ex);
        \foreach \y in {-0.21,0.19}
        {
            \node at (0,\y) {\scriptsize $\cdots$};
        }
        \node at (-1.1,0.6) {\scriptsize $K$};
        \node at (1.1,0.6) {\scriptsize $L$};
        \node at (-1.1,-0.6) {\scriptsize $K'$};
        \node at (1.1,-0.6) {\scriptsize $L'$};
    \end{tikzpicture}
}
\def\tikzTwoLegs{
    \begin{tikzpicture}[baseline=-3pt]
    \draw (0,-.6) -- (0,.6);
    \draw (.5,-.6) -- (.5,.6);
    \node at (0,-.8) {\scriptsize $\sigma_1$};
    \node at (.5,-.8) {\scriptsize $\sigma_{N-1}$};
    \node at (0,.8) {\scriptsize $\lambda_1$};
    \node at (.5,.8) {\scriptsize $\lambda_{N-1}$};
    \end{tikzpicture}
}

\begin{titlepage}
\begin{center}
\vspace{2cm}

{\fontsize{20.5pt}{25pt}\selectfont
Bootstrapping Open Quantum Many-body Systems\\with Absorbing Phase Transitions
}

\vspace{2cm}

Minjae Cho$^{1,\dagger}$, 
Colin Oscar Nancarrow$^{2,\ddagger}$, 
Petar Tadi\'c$^{3,4,\S}$, 
Yuan Xin$^{5,6,\P}$

\vspace{1cm}

{\it
$^1$Leinweber Institute for Theoretical Physics, University of Chicago, Chicago, IL 60637, USA\\[0.2cm]
$^2$Department of Physics, Boston University, Boston, MA 02215, USA\\[0.2cm]
$^3$Department of Physics, University of Oxford, Oxford, OX1 3PU, United Kingdom\\[0.2cm]
$^4$Institute for Interdisciplinary and Multidisciplinary Studies, University of\\
Montenegro, 81000 Podgorica, Montenegro\\[0.2cm]
$^5$Center for Mathematics and Interdisciplinary Sciences, Fudan University,\\ Shanghai, 200433, China\\[0.15cm]
$^6$Shanghai Institute for Mathematics and Interdisciplinary Sciences (SIMIS),\\ Shanghai, 200433, China
}

\vspace{1cm}

\end{center}

\noindent
We demonstrate that combining the positivity of density matrices with steady-state conditions yields a systematic bootstrap method for studying open quantum many-body systems governed by Lindblad master equations on infinite lattices, which exhibit absorbing phase transitions. As a concrete example, we apply this method to the quantum contact process with an absorbing state. We obtain bootstrap bounds on steady-state expectation values, the critical coupling, certain ratios of expectation values in the nontrivial steady state in the supercritical phase, and the Liouvillian spectral gap in the subcritical phase.

\blfootnote{$^\dagger$\texttt{cho7@uchicago.edu}}
\blfootnote{$^\ddagger$\texttt{con@bu.edu}}
\blfootnote{$^\S$\texttt{petar.tadic@physics.ox.ac.uk}}
\blfootnote{$^\P$\texttt{yuan\_xin@fudan.edu.cn}}

\end{titlepage}

\tableofcontents

\section{Introduction}

The study of closed quantum systems described by Hamiltonians has been one of the main driving forces in the development of physics over the past century. In particular, the program of classifying phases of quantum many-body systems has achieved great success and continues to broaden its scope.

An intriguing extension that has attracted increasing attention in recent years concerns the phases of open quantum many-body systems. Due to coupling with an external environment, such systems undergo quantum dissipation, and their time evolution is no longer unitary. Nevertheless, as system parameters are varied, they may exhibit phase transitions characterized by macroscopic or information-theoretic order parameters (see e.g. \cite{Casteels:2017ecv,Minganti:2018kgs,Skinner:2018tjl,Li:2018mcv,Lee:2023fsk} for a few examples). These transitions are fundamentally distinct from those in closed systems and play a crucial role in understanding quantum decoherence, which arises in both theoretical models and experimental platforms.

The prototypical universality class of such nonequilibrium phase transitions is the directed percolation (DP) universality class, whose classical models have been extensively studied in the past (see \cite{Hinrichsen:2000fz} for a comprehensive review) and whose quantum extensions have been discussed in recent years \cite{2016PhRvL.116x5701M,Carollo:2019zmt,Gillman:2019lfe,Jo:2021tax}. One of its characteristic features is the existence of an absorbing state: a pure state that remains a steady state both below and above criticality. The absorbing state is the only steady state in the subcritical phase, while other genuinely mixed steady states appear in the supercritical phase. Despite being a prototypical example, this class has not been solved analytically even in one dimension, remaining an outstanding challenge in mathematical and theoretical physics.

In general, rigorous theoretical results on open quantum many-body systems are much harder to find than those on closed systems. The absence of a Hermitian time-evolution generator and of symmetries that support the Landau paradigm is both an outstanding feature and a practical obstacle, leaving relatively few available methods for analyzing important physical observables.

In this work, we demonstrate that a bootstrap method based on the positivity of density matrices provides a systematic framework for deriving rigorous results for open quantum many-body systems on infinite lattices governed by Lindblad master equations and exhibiting absorbing phase transitions. Similar bootstrap techniques have previously been applied to equilibrium states of closed quantum systems \cite{Barthel:2012mqo, Han:2020bkb, Nancarrow:2022wdr, Fawzi:2023fpg, Gao:2024etm, Cho:2024kxn, Cho:2024owx, Cho:2025vws} and, more recently, to steady states of Lindbladian dynamics  \cite{Robichon:2024apx, Mortimer:2024fuu, Cho:2025nlv}. Related approaches have also been successfully applied to classical equilibrium and nonequilibrium systems \cite{Anderson:2016rcw, Anderson:2018xuq,Kazakov:2021lel,Kazakov:2022xuh,Cho:2022lcj,Cho:2023ulr, Cho:2025dgc}.

A recent development particularly relevant to this work is a bootstrap framework for nonequilibrium stochastic processes on infinite lattices, in which the positivity of probability measures and master equations were combined to yield rigorous and sharp bounds on observables in systems with absorbing states \cite{Cho:2025dgc}. The present work extends this framework to quantum systems described by Lindblad master equations, focusing in particular on systems with absorbing phase transitions. Concretely, we obtain bootstrap bounds on steady-state expectation values that lead to lower bounds on the critical coupling, and we further formulate new bootstrap constraints tailored to the nontrivial steady state in the supercritical phase and to the Liouvillian spectral gap in the subcritical phase.

Steady states are represented by density matrices that remain invariant under Lindblad time evolution. Consequently, they must satisfy two key properties: (1) positivity of the density matrix, and (2) stationarity under the Lindblad dynamics. In the bootstrap method, these defining properties are treated as constraints on the full space of density matrices. For quantum many-body systems on an infinite lattice, they give rise to infinitely many such constraints. The key idea is that even a finite subset can yield rigorous bounds on physical observables, since any true steady state must satisfy all of them. As more constraints are imposed, these bounds are expected to become progressively sharper.

Before turning to the specific bootstrap formulation for the Lindblad master equation, we emphasize that the idea of constraining the physics by imposing its defining properties has appeared across many areas of mathematical science. Examples include quantum chemistry \cite{PhysRevA.57.4219, 10.1063/1.1360199}, conformal bootstrap \cite{Rattazzi:2008pe,Rattazzi:2010gj,Rattazzi:2010yc}, dynamical systems and stochastic processes \cite{hernandez2012markov,doi:10.1137/15M1053347,2018PhLA..382..382T,2018arXiv180708956K}, and optimal control theory \cite{2007math......3377L,2022arXiv221115652H}.

\subsection{Setup}

\noindent The Lindblad master equation \cite{Lindblad:1975ef,Gorini:1975nb} is the most general Markovian, trace-preserving, completely positive evolution for a density matrix $\rho$ (see e.g. \cite{Manzano:2020yyw} for a brief introduction to the subject)
\begin{equation}
    \frac{d\rho}{dt}= \mathcal{L}(\rho)\equiv -i[H,\rho] +  \sum_{k\in{\cal W}} \gamma_{(k)}\left(L^{(k)\dagger}\rho L^{(k)}-\frac{1}{2}\{L^{(k)\dagger}L^{(k)},\rho \} \right) ,
\end{equation}
where $H$ is the Hamiltonian of the system, $\cal W$ is an appropriate index set, $L^{(k)}$ are Lindblad jump operators that parametrize the effective system-environment interaction after tracing out the environment, and $\gamma_{(k)}\geq0$ are the corresponding rates. The density matrix is Hermitian, unit-normalized and positive semidefinite
\begin{equation}\label{rhoconstraints}
    \rho^{\dagger} = \rho,\qquad {\rm Tr}(\rho)  = 1,\qquad \rho \succeq 0.
\end{equation}
Steady states $\rho$ are defined by $\frac{d\rho}{dt} = 0$. The number and structure of such steady states of a Lindblad dynamics are determined by the kernel of the Lindbladian $\mathcal{L}$
\begin{equation}
    {\rm ker}\,\mathcal{L} = \{\rho \,\,|\,\, \mathcal{L}(\rho) = 0\},
\end{equation}
which depends on the Hamiltonian and jump operators. As system parameters in the Hamiltonian and $\gamma_{(k)}$ are varied, the dimension of this kernel can change, signaling the phase transition. One possibility for such a change in the dimension of the kernel is the closing of the spectral gap.

In this work, we focus on the quantum spin system on the one-dimensional lattice whose Lindblad equation in the Heisenberg picture takes the form
\begin{equation}\label{master}
    \frac{d}{dt} \langle \CO\rangle = \langle\hat{\CL}(\CO)\rangle_\rho \equiv i \langle [H, \CO] \rangle_\rho + \gamma\sum_{k\in \mathbb{Z}} \left( \left\langle L^{(k)\dagger} \CO L^{(k)} \right\rangle_\rho
    -\frac{1}{2} \left\langle \left\{ L^{(k)\dagger} L^{(k)}, \CO \right\} \right\rangle_\rho \right).
\end{equation}
Note that on each lattice site $k$, we have a single jump operator $L^{(k)}$ whose corresponding rate $\gamma^{(k)}=\gamma$, independent of the site. The bootstrap method implements~\eqref{rhoconstraints},~\eqref{master} and the translation invariance of the system as the constraints on the space of density matrices and produces rigorous bounds on their expectation values.

It is worth mentioning that equations obeyed by expectation values such as \eqref{master} can be taken as the \textit{definition} of the system on the \textit{infinite} lattice. Bootstrap formulation implemented in this work only imposes constraints on the space of such expectation values. Even though there are a priori infinitely many bootstrap constraints for such systems on the infinite lattices, we in practice implement only a finite subset of such constraints. The key point is that those finite subsets of constraints are still supposed to hold for the infinite lattice systems. Therefore, bootstrap bounds presented in this work apply to the systems on the infinite lattice.

\subsection{Quantum contact process}

\noindent The main example of Lindblad dynamics we investigate in this work is the quantum contact process. It provides an open quantum generalization of the classical contact process \cite{10.1214/aop/1176996493, liggett2004interacting}, a paradigmatic example of nonequilibrium phase transitions in the directed percolation universality class \cite{Hinrichsen:2000fz}.

We consider the one-dimensional chain where each site carries a qubit with basis states
\begin{equation}
{\rm inactive}:\,\,\,    |0\rangle = \begin{pmatrix}
                    0 \\
                    1 
                \end{pmatrix}, \qquad {\rm active:}\,\,\,    |1\rangle = \begin{pmatrix}
                    1 \\
                    0 
                \end{pmatrix}. 
\end{equation}  
The Hamiltonian is given by
\begin{equation}
    H = \Omega \sum_{k\in \mathbb{Z}} \left( \sigma_1^{(k)} n^{(k+1)}+ n^{(k)}\sigma_1^{(k+1)} \right), 
\end{equation}
where $\Omega>0$ is the strength of the Hamiltonian part of the dynamics,
\begin{equation}
    n^{(k)}=\sigma^{(k)}_{+}\sigma^{(k)}_{-}, \qquad \sigma^{(k)}_{\pm}=\frac{1}{2}\left(\sigma_{1}^{(k)}\pm i\sigma_{2}^{(k)}\right),
\end{equation}
with $\sigma_{1,2,3}$ being the Pauli matrices, and the jump operators are
\begin{equation}
    L^{(k)} = \sigma^{(k)}_{-},
\end{equation}
at each lattice site $k$. Without loss of generality, we fix $\gamma^{(k)}=\gamma=1$.

It is straightforward to verify that the the density matrix
\begin{equation}
    \rho_0=\left(\underset{k\in \mathbb{Z}}\otimes|0\rangle^{(k)}\right)\left(\underset{k\in \mathbb{Z}}\otimes\langle0|^{(k)}\right),
\end{equation}
is a pure steady state of the system, which is called the absorbing state. Indeed, $\rho_0$ is annihilated by all jump operators, and since the Hamiltonian contains only terms proportional to the local occupation operators, it also annihilates $\rho_0$. Consequently, all contributions to the Lindblad equation vanish, and $\rho_0$ is a steady state. 

Depending on whether the value of Hamiltonian coupling $\Omega$ is below or above the critical value $\Omega_*$, the system can be in one of the two phases: 
\begin{itemize}
    \item {\bf Absorbing phase ($\Omega\leq\Omega_*$)}: the state  $\rho_0$ is the unique steady state.

    \item {\bf Active phase ($\Omega>\Omega_*$)}: in addition to the absorbing state $\rho_0$, the system admits other nontrivial steady states.
\end{itemize}
In the absorbing phase, any initial state $\rho(0)$ relaxes to $\rho_0$ at late times with an exponential tail, whose exponent $\Delta$ is identified with the Liouvillian spectral gap. In the active phase, it is expected that there is a unique nontrivial translation-invariant steady state $\rho_1$ that maximizes the magnetization $\langle Z_1 \rangle$, where $Z_k\equiv \sigma_3^{(k)}$, and all the other translation-invariant steady states are given by the weighted averages of $\rho_0$ and $\rho_1$.

By imposing the constraints coming from~\eqref{rhoconstraints},~\eqref{master} and the translation invariance of the steady states, we proceed to obtain the upper and lower bounds on the expectation values of the magnetization $\langle Z_1 \rangle$ together with the lower bound on critical coupling $\Omega_*$ (section \ref{sec:steadyDirect}), upper and lower bounds on the ratio of deviations $\delta \langle Z_1 Z_2 \rangle/\delta\langle Z_1 \rangle$, where $\delta\langle \mathcal{O} \rangle = \langle\mathcal{O} \rangle_{\rho_1}-\langle\mathcal{O} \rangle_{\rho_0}$ (section \ref{sec:ratio}), and the Liouvillian spectral gap $\Delta$ (section \ref{sec:gap}).

\section{Bootstrapping steady states}\label{sec:steadyDirect}

\noindent In this section we set up the bootstrap problem for the steady state. We begin with relation~\eqref{master}. For a steady state $\rho$, expectation values do not evolve in time
\begin{equation}\label{steady-state}
    \frac{d}{dt} \langle \CO\rangle_{\rho} = \langle\hat{\CL}(\CO)\rangle_\rho = 0,
\end{equation}
where $\hat{\CL}$ is defined in~\eqref{master}. We use the steady state condition \eqref{steady-state} as the basic equations of motion. The equations are local and only maps a local observable with a support on $\ell$ sites to observables with a slightly larger support on $(\ell+2)$ sites. Thus we can conveniently set a truncation scheme were we restrict to the observables with support on at most $N$ sites. The search space runs over the entire local Hilbert space of $N$ sites:
\begin{equation}
    \mathcal{S}_{N} =\left\{ \left. \prod_{k=1}^{N}  \sigma_{s_k}^{(k)}   \right| s_k = 0,1,2,3 \right\},
\end{equation}
where $\sigma_{0}\equiv \mathbbm{1}_{2\times 2}$. Any product of operators in this set falls back to the same set $\mathcal{S}_{N} \times \mathcal{S}_{N} \rightarrow \mathcal{S}_{N}$. This space has a naive dimension of $4^N$. The equation of motion, i.e. steady-state condition~\eqref{steady-state} serves as linear constraints on $\langle \mathcal{S}_{N}\rangle$. We further assume that the steady state is translation-invariant, meaning that a fixed string of operators will have the same expectation value independent of the starting position of the string. The condition
\begin{equation}\label{translation-inv}
    \left\langle \mathcal{T}(\CO)\right\rangle_{\rho} = \left\langle \CO\right\rangle_{\rho}, \qquad
    \mathcal{T}\left( \mathbbm{1}^{(1)}\prod_{k=2}^{N}  \sigma_{s_k}^{(k)}\right) \equiv 
    \left(\prod_{k=2}^{N}  \sigma_{s_k}^{(k-1)}\right)\mathbbm{1}^{(N)}, 
\end{equation}
covers all translation invariance relations within $\mathcal{S}_N$. We add the one site bleed on the edge because the Hamiltonian acts on $\mathcal{S}_{N-1}$ and naively maps it to $\mathcal{S}_{N+1}$, but the bleed in translation invariance will map it back to $\mathcal{S}_{N}$. The steady-state bootstrap for the subsystem size $N$ can be summarized as the following
\begin{equation}\label{steady-state-setup}
\begin{array}{rll}
    \text{minimize/maximize}~ &\langle \CO_{\rm obj} \rangle_{\rho}, & \\
    \text{in parameter space}~ & \langle\CO\rangle_{\rho}~~\text{for}~\CO\in{\cal S}_N, & \\
    \text{subject to}~
    & M_{ij} \equiv \langle \CO_i^\dagger \CO_j \rangle_{\rho} \succeq 0~~ \forall \CO_{i,j}\in \mathcal{S}_N, & \\
    & \langle \mathcal{T}(\CO) \rangle_{\rho} = \langle \CO \rangle_{\rho}, \quad \forall \CO \in \mathcal{S}_{N-1}, & \quad (\ref{translation-inv})\\
    & \langle \hat{\CL}(\CO)\rangle_\rho = 0,~~ \forall \CO\in \mathcal{S}_N, & \quad (\ref{steady-state}) \\ 
    & \langle \mathbbm{1} \rangle_{\rho} = 1. 
\end{array}
\end{equation}
The obtained minimum and maximum value of $\langle \CO_{\text{obj}}\rangle_\rho$ respectively provide rigorous lower and upper bounds on the steady-state value of $\langle \CO_{\text{obj}}\rangle_\rho$ for the system on the infinite lattice.

\subsection{Speed-up with reduced density matrix}
The setup (\ref{steady-state-setup}) can be further optimized by choosing a natural basis that corresponds to the reduced density matrix. We define the basis operators to be all operators of the shape
\begin{equation}
    | \nu_1 \nu_2 \cdots \nu_k \rangle \langle \mu_1 \mu_2 \cdots \mu_k |
    = \cdots \mathbbm{1}^{(0)}\otimes | \nu_1 \rangle \langle \mu_1 | \otimes | \nu_2 \rangle \langle \mu_2 | \otimes | \nu_k \rangle \langle \mu_k | \otimes \mathbbm{1}^{(k+1)} \cdots,
\end{equation}
for $k=1,...,N$, which is equivalent to the original local basis
\begin{equation}
    \mathcal{S}_k = \bigg\{ | \nu_1 \nu_2 \cdots \nu_k \rangle \langle \mu_1 \mu_2 \cdots \mu_k | ~\bigg|~ \mu_i,\nu_i \in \{0,1\} \bigg\}~.
\end{equation}
The matrix of expectation values of the operators is the reduced density matrix $\rho^{(k)}$ of consecutive $k$ sites
\begin{equation}
    \big\langle | \nu_1 \nu_2 \cdots \nu_k \rangle \langle \mu_1 \mu_2 \cdots \mu_k | \big\rangle_\rho = \big(\rho^{(k)}\big)_{\nu_1 \nu_2 \cdots \nu_k}^{\mu_1 \mu_2 \cdots \mu_k},
\end{equation}
where we call $k$ the level of the reduced density matrix. There is a hierarchy of constraints \cite{Kull:2022wof} between reduced density matrices of nearby levels,
\begin{equation}\label{eqn:RTI}
    \rho^{(k)} = {\rm Tr}_1 \rho^{(k+1)} = {\rm Tr}_{k+1}\rho^{(k+1)},
\end{equation}
which also captures the translation invariance condition (\ref{translation-inv}).
The Liouvillian super-operator action (\ref{master}) can also be written as a hierarchy of constraints between reduced density matrices of nearby levels following the procedure in \cite{Cho:2024owx}.

Moreover, the matrix $M$ in \eqref{steady-state-setup} can be factorized as
\begin{equation}
    M_{ij} = \langle \psi_i | \phi_j\rangle \langle\psi_j | \rho^{(N)} | \phi_i\rangle,
\end{equation}
where $\psi_i$ and $\phi_j$ are basis states $| \nu_1 \nu_2 \cdots \nu_N \rangle$. Therefore, the positive-semidefiniteness of $M$ is equivalent to
\begin{equation}
    \rho^{(N)} \succeq 0.
\end{equation}
Note that this implies $\rho^{(k)}\succeq0$ for $k=1,\cdots,N-1$ due to \eqref{eqn:RTI}. The dimension of the positivity matrix now reduces from $4^N$ to $2^N$, allowing the semidefinite programming procedures to be conducted at higher $N$. The same basis is used in all of the following sections.

\subsection{Bootstrap results}
\begin{figure}[h!]
    \centering
    \includegraphics[width=0.48\linewidth]{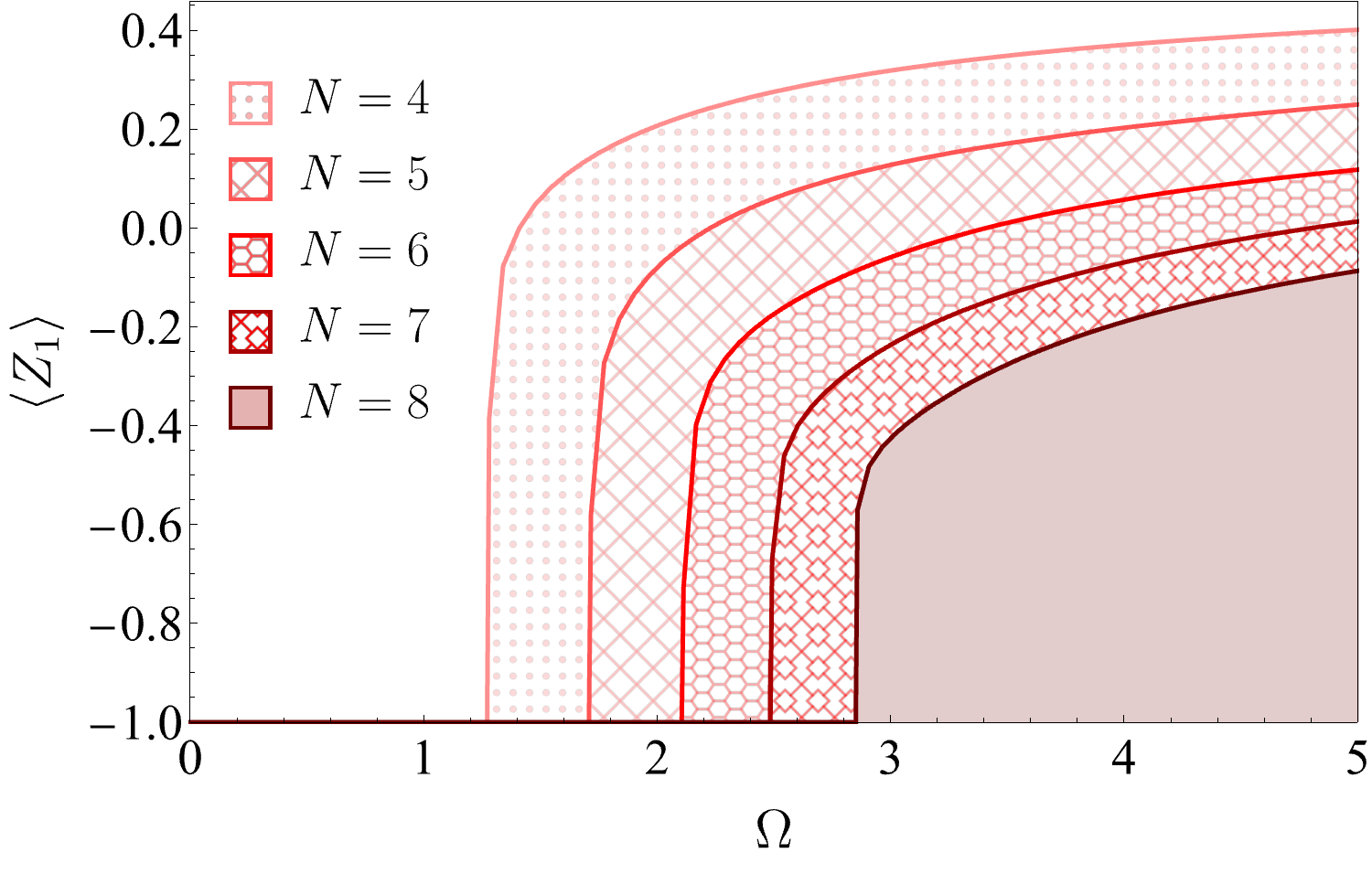}
    \includegraphics[width=0.48\linewidth]{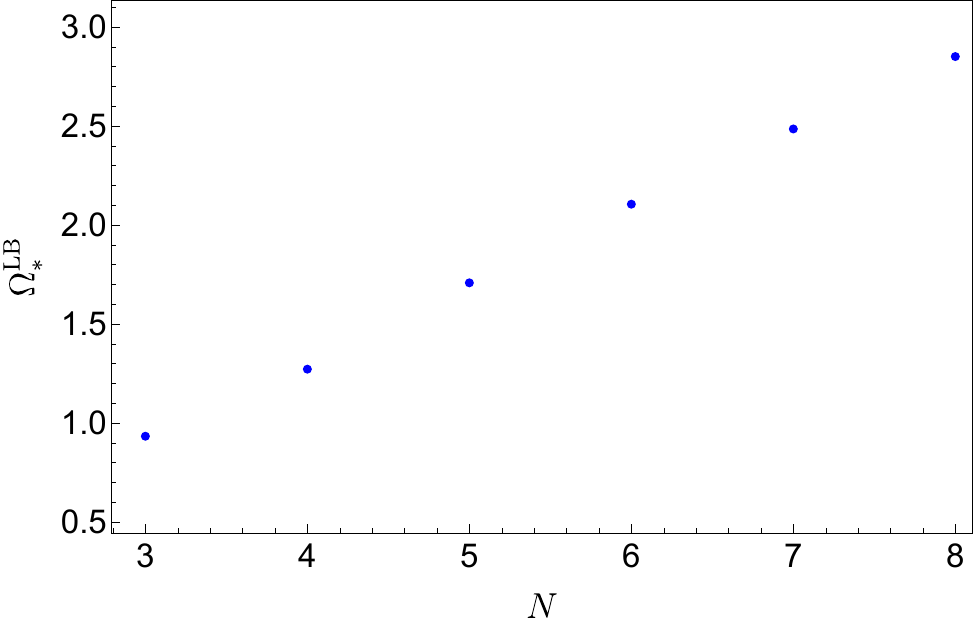}
    \caption{Left: The bounds on $\langle Z_1\rangle$ in the steady state as functions of the coupling $\Omega$ for various subsystem sizes $N$. Right: The dependence of the lower bound on critical value of coupling $\Omega_{*}$ on the subsystem size $N$.}
    \label{fig:magnetization&OS}
\end{figure}

We start by bootstrapping the bounds on the expectation values of magnetization $Z_k\equiv \sigma_3^{(k)}$ in the steady state as the function of coupling $\Omega$. The results are shown in fig.~\ref{fig:magnetization&OS}. For low enough value of $\Omega$, the bootstrap upper bound on $\langle Z_1\rangle$ coincides with $-1$. Such values of $\Omega$ then provides a rigorous lower bound on the critical value of the coupling $\Omega_*$. For $\Omega$ greater than this value, the bootstrap constraints imposed at the subsystem size $N$ do not exclude the existence of the non-trivial steady state where magnetization satisfies $\langle Z_1 \rangle > -1$. 

The maximum value of $N$ we implemented was $N=8$, where we find the lower bound on $\Omega_*$ to be
\begin{equation}
    \Omega_*^{\rm LB}|_{N=8} = 2.850755.
\end{equation}
From the right plot in fig.~\ref{fig:magnetization&OS}, it is clear that the lower bounds on $\Omega_*$ for $N\leq8$ have not yet converged and that one should be able to improve them further by imposing the constraints at higher $N$. As a reference, \cite{Carollo:2019zmt} gave an estimate $\Omega_*\approx6$ based on the infinite time-evolving
block decimation algorithm.

\begin{figure}[h!]
    \centering
    \includegraphics[width=0.90\linewidth]{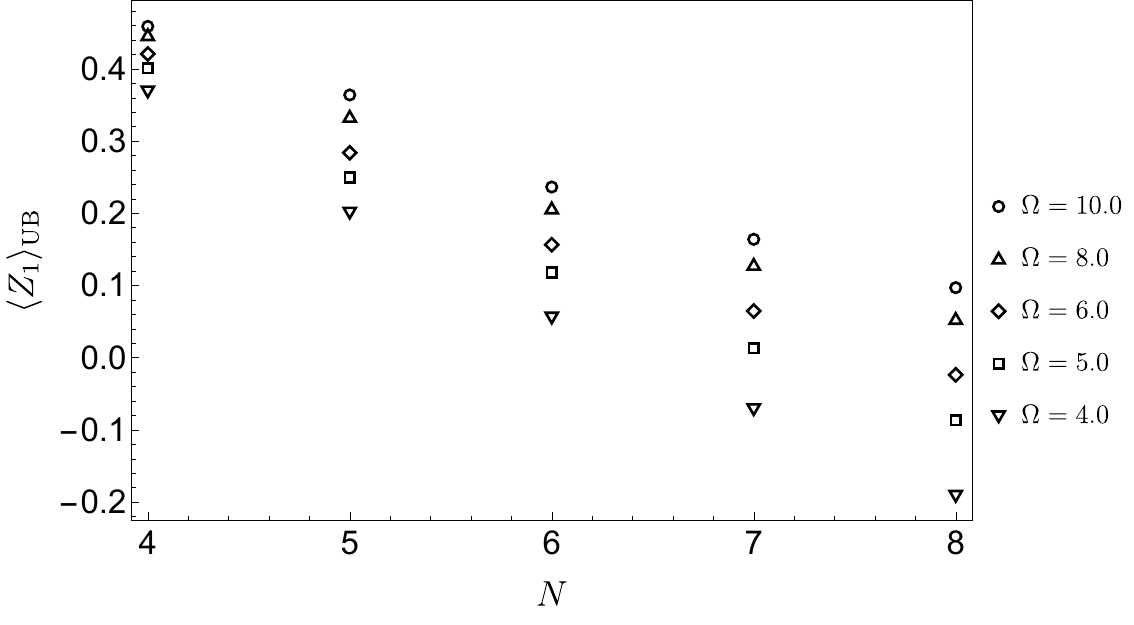}
    \caption{The dependence of the upper bounds on $\langle Z_1\rangle$ in the steady state on the subsystem size $N$ for different values of coupling $\Omega$. }
    \label{fig:ZUB}
\end{figure}

In fig.~\ref{fig:ZUB}, we show the dependence of the upper bound on magnetization on the subsystem size $N$ for various values of coupling $\Omega$ that are all greater than $\Omega^{\rm LB}_{*}|_{N=8}$. It is again clear that the upper bounds have not converged yet with subsystem sizes $N\leqslant 8$, so larger values of $N$ are expected to improve the results significantly.

\section{Bootstrapping the nontrivial steady state}\label{sec:ratio}

\noindent In this section, we obtain bootstrap upper and lower bounds on the ratio of deviations of $Z_1 Z_2$ and $Z_1$, to be defined below, that a priori exists only for $\Omega>\Omega_*$. Assuming that we are in such an active phase, we start by considering the following difference between density matrices
\begin{equation}
    D = \rho_1 - \rho_0,
\end{equation}
where $\rho_1$ is the non-trivial steady state maximizing $\langle Z_1\rangle$ in the active phase. Then, all steady states can be written as
\begin{equation}
    \rho_{w}=\rho_0 + w D, \qquad w\in [0,1].
\end{equation}
We define the deviation of observables as the expectation value in a priori unphysical state $D$ to measure how much the observables in the non-trivial steady state deviate from the absorbing state
\begin{equation}
    \delta\langle \CO \rangle \equiv \langle \CO \rangle_{D} =  \langle \CO \rangle_{\rho_1}-\langle \CO \rangle_{\rho_0}.
\end{equation} 

It is straightforward to verify that $D$, when viewed as a matrix, is positive semidefinite in the kernel space of $\rho_0$ due to positive-semidefiniteness of $\rho_1$. For the subsystem size $N$, we consider the matrix elements of $\rho_0$ and $D$ denoted by  $(\rho_0^{(N)})_{ij}=\langle\CO^\dagger_i\CO_j\rangle_{\rho_0}$ and $(D^{(N)})_{ij} = \langle\CO^\dagger_i\CO_j\rangle_{D}$ built from $\CO_{i,j}\in\mathcal{S}_N$. We obtain the basis of $\rho_0^{(N)}$ null vectors by solving $\rho_0^{(N)}b^\sigma=0$. We denote by $\Sigma_N$ the index set such that $b^\sigma$ for $\sigma\in\Sigma_N$ provides a basis for such null vectors. Matrix $D^{(N)}$ can be projected to the null space of $\rho_0^{(N)}$ as $\left(\mathcal{D}^{(N)}\right)^{\sigma\sigma'} \equiv (b^{\sigma}_{i})^{*}(D^{(N)})_{ij} b^{\sigma'}_{j}, $ where the $\mathcal{D}^{(N)}$ should satisfy $\mathcal{D}^{(N)}\succeq0$.


Furthermore, since both $\rho_0$ and $\rho_1$ are steady states, the steady state conditions~\eqref{steady-state} are also satisfied by the expectation values in $D$. However, one key distinction of $\langle\cdots\rangle_D$ from $\langle\cdots\rangle_\rho$ is that $\langle \mathbbm{1} \rangle_{D} = 0$, thereby making $\mathcal{D}^{(N)}\succeq0$ and \eqref{steady-state} for $\langle\cdots\rangle_D$ linear and homogeneous constraints on $\langle\CO\rangle_D$'s. Therefore, these bootstrap constraints can not differentiate between $D$ and $\alpha D$ for any $\alpha>0$. To fix this scale $\alpha$ in the constraints, we define a {\it ratio of deviations} 
\begin{equation}\label{rdef}
    r(\CO) \equiv\frac{\delta \langle \CO \rangle}{\delta \langle Z_1 \rangle} = \frac{\langle \CO \rangle_{\rho_1} - \langle \CO \rangle_{\rho_0}}{\langle Z_1 \rangle_{\rho_1} - \langle Z_1 \rangle_{\rho_0}},
\end{equation}
which will become bootstrap variables below. In the active phase, $\delta \langle Z_1 \rangle > 0$, so $r(\mathcal{O})$ is well-defined. Furthermore, defining
\begin{equation}
({\cal R}^{(N)})^{\sigma\sigma'}\equiv(b^{\sigma}_{i})^{*}~r(\CO^\dagger_i\CO_j)~b^{\sigma'}_{j}={({\cal D}^{(N)})^{\sigma\sigma'}\over \delta\langle Z_1\rangle},
\end{equation}
we obtain ${\cal R}^{(N)}\succeq0$ as a result of ${\cal D}^{(N)}\succeq0$. Also, it is easy to see that $r(\mathbbm{1})=0$ and $r(Z_1)=1$, the latter of which makes the bootstrap problem inhomogenous.



Now, one can set up the bootstrap problem for variables $r(\CO)$ as follows
\begin{equation}
\begin{array}{rll}
    \text{minimize/maximize}~ & r(\CO_{\text{obj}}), \\
    \text{in parameter space}~ & r(\CO)~~\text{for}~\CO\in{\cal S}_N, \\
    \text{subject to}~
    & (\mathcal{R}^{(N)})^{\sigma\sigma'} \succeq 0, \\
    & r(\mathcal{T}(\CO)) = r(\CO), \quad \forall \CO \in \mathcal{S}_{N-1}, \\
    & r(\hat{\CL}(\CO)) = 0,~~ \forall \CO\in \mathcal{S}_N,  \\ 
    & r(\mathbbm{1})=0, \\
    & r(Z_1)=1.
\end{array}
\end{equation}
Even though $Z_1$ provides the simplest choice for the reference point in defining the ratios $r(\CO)$, one may choose other observables as a reference point. Since we already obtained nontrivial bounds on $\langle Z_1\rangle_\rho$ for steady states, bounds on the above $r(\CO_{\text{obj}})$ may further be turned into bounds on $\langle\CO_{\text{obj}}\rangle_{\rho_1}$ itself if one wishes.

\begin{figure}
    \centering
    \includegraphics[width=1.00\linewidth]{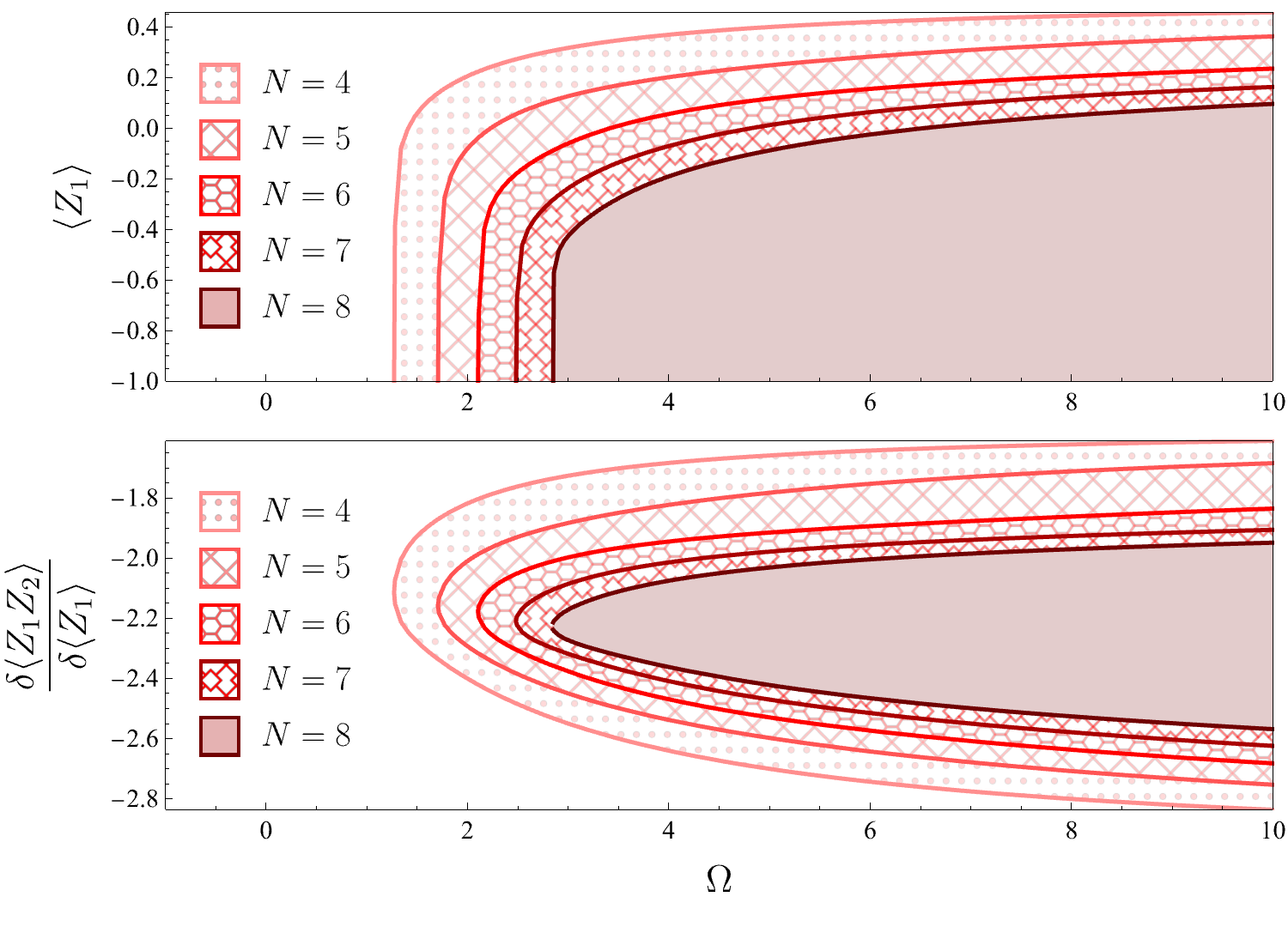}
    \caption{Upper: the upper bound on $\langle Z_1\rangle$ for steady states as a function of $\Omega$. Lower: the upper and lower bounds on $\delta \langle Z_1 Z_2\rangle/\delta\langle Z_1\rangle$ as a function of $\Omega$. }
    \label{fig:ZandRatio}
\end{figure}

We obtain the bootstrap upper and lower bounds on $r(Z_1Z_2)$. From this bootstrap problem, one can again extract the lower bounds on the critical coupling $\Omega_*$ by finding the lowest value of $\Omega$ for which the constraints have a solution. The values of lower bounds on $\Omega_*$ agree with those obtained from the steady-state bootstrap performed in section \ref{sec:steadyDirect}, confirming the consistency of the methods. We show both the steady-state bounds on $\langle Z_1\rangle$ and bounds on $r(Z_1Z_2)$ as functions of $\Omega$ in fig.~\ref{fig:ZandRatio}.

\begin{figure}
    \centering
    \includegraphics[width=1.00\linewidth]{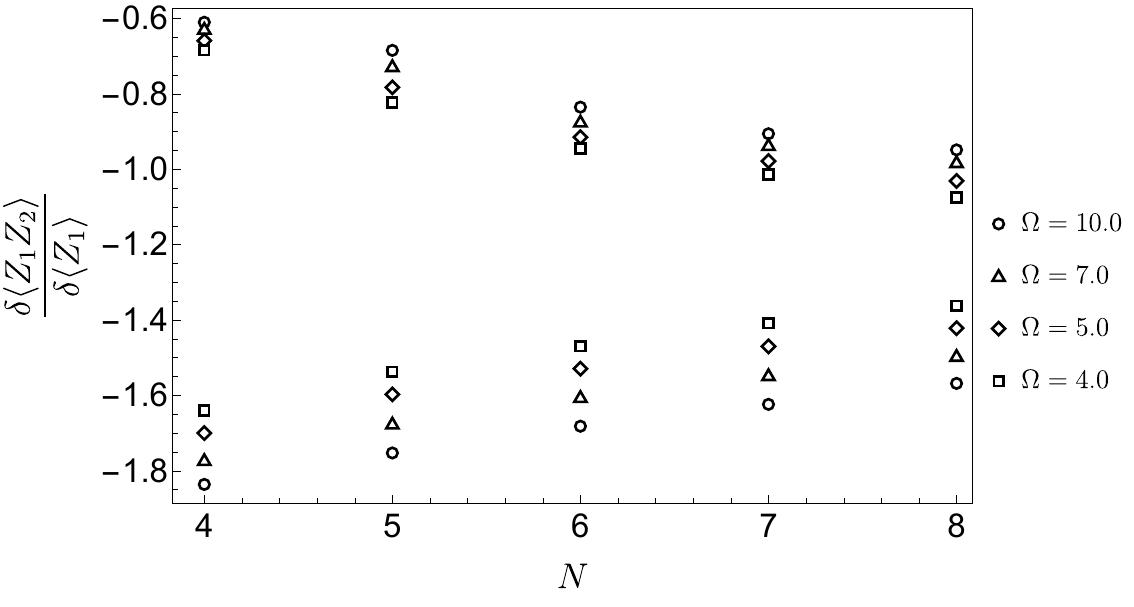}
    \caption{The dependence of the upper and lower bounds on $r$-observable \eqref{rdef} with $\mathcal{O}_{\rm obj} = Z_1 Z_2$ and $\mathcal{O}_{\rm ref} = Z_1$ on the subsystem size $N$ for different values of $\Omega$. }
    \label{fig:RatioULB}
\end{figure}

We also show the upper and lower bounds on $r(Z_1Z_2)$ as functions of the subsystem size $N$ for various values of coupling ($\Omega>\Omega_*^{\rm LB}$) in fig.~\ref{fig:RatioULB}. Similarly to steady-state case in section \ref{sec:steadyDirect}, one can see that the bounds have not yet converged for the subsystem sizes $N\leqslant 8$ and expect nontrivial improvements at higher $N$.

\section{Bootstrapping the Liouvillian spectral gap}\label{sec:gap}

\noindent In this section, we obtain the upper and lower bounds on the Liouvillian spectral gap $\Delta$ in the absorbing phase. In this phase, any initial state evolves towards the absorbing state $\rho_0$. Correspondingly, the expectation values relax exponentially towards their absorbing state values at late times:
\begin{equation}\label{eqn:latetimedecay}
    \langle \CO \rangle(t) \rightarrow \langle \CO \rangle_{\rho_0} + B_{\CO} e^{-\Delta t} + \ldots,~~~\text{as }t\rightarrow\infty,
\end{equation}
where $B_{\CO}$ encodes the contribution of the leading decay mode. Importantly, the value of $\Delta$ is universal for all the operators $\CO$. Substituting the late-time behavior \eqref{eqn:latetimedecay} into the Lindblad equation~\eqref{master} yields a closed set of linear equations for the coefficients $B_{\CO}$ that depends on $\Delta$:
\begin{equation}\label{eqn:latetimeevo}
    -\Delta B_{\CO} e^{-\Delta t} = \langle \hat{\CL}(\CO)\rangle_{\rho_0}+B_{\hat{\CL}(\CO)}e^{-\Delta t}~~~\Rightarrow~~~-\Delta B_{\CO} = B_{\hat{\CL}(\CO)}.
\end{equation}
Here, we used that $\langle \hat{\CL}(\CO)\rangle_{\rho_0}=0$ since the absorbing state is a steady state. Note that $B_{\CO}$ is linear in $\CO$, in that if $\CO=\sum_\alpha \CO_\alpha$ for some basis of operators $\CO_\alpha$, then $B_{\CO}=\sum_\alpha B_{\CO_\alpha}$. Therefore, for a fixed value of $\Delta$, the late-time equations of motion \eqref{eqn:latetimeevo} define a homogeneous linear system in $B_{\CO}$.

To impose positivity, we consider moment matrices at late times
\begin{equation}
    M_{ij}(t) = \langle \CO_{i}^{\dagger}\CO_j \rangle (t) \rightarrow (\rho_0)_{ij} + \mathcal{M}_{ij}e^{-\Delta t},~~~\text{as }~t\rightarrow\infty,
\end{equation}
where $\mathcal{M}_{ij}=B_{\CO_{i}^{\dagger}\CO_j}$. Since $M(t)\succeq 0$ for all $t$, it follows that
\begin{equation}
    \mathcal{M}\succeq 0 \quad \text{on the null space of } \rho_0.
\end{equation}
The moment matrix projected onto the null space $\mathcal{B}^{\sigma\sigma'} \equiv (b_i^\sigma)^* \mathcal{M}_{ij} b_j^{\sigma'}$ provides a matrix that should be positive semidefinite on its own. Note that $\mathcal{B}$ is a matrix linear in $B_\CO$ and its positivity therefore provides a set of linear matrix inequalities on $B_{\CO}$.
We define $\mathbbm{1}_{\Sigma_N}$ to be the identity matrix in the space of $\sigma$ indices in $\Sigma_N$: $(\mathbbm{1}_{\Sigma_N})^{\sigma\sigma'}=\delta^{\sigma\sigma'}$ for $\sigma,\sigma'\in\Sigma_N$.

Because the all of the above constraints on are homogeneous in $B_{\CO}$, we fix the overall scale by imposing a normalization condition, $B_{Z_1} = 1$. For a given test value of $\Delta$, we then solve a feasibility problem:
\begin{equation}\label{gap-search}
\begin{array}{rll}
    \text{minimize}~ & g,  \\
    \text{in parameter space}~ & g~\text{and}~B_\CO,~~\text{for}~ \CO\in\mathcal{S}_N\\ 
    \text{subject to}~
    &  -\Delta B_{\CO} = B_{\hat{\CL}(\CO)},~~ \forall \CO \in \mathcal{S}_{N}, \\ 
    & B_{\mathcal{T}(\CO)} = B_\CO, \quad \forall \CO \in \mathcal{S}_{N-1}, \\
    & B_{Z_1} = 1, \\
    & \mathcal{B}^{(N)} + g \mathbbm{1}_{\Sigma_N} \succeq 0. 
\end{array}
\end{equation}
If the minimal value satisfies $g_{\rm min}>0$, then the value of $\Delta$ is not feasible. Conversely, if $g_{\rm min}\leqslant0$, then a feasible solution exists and $\Delta$ is allowed. By scanning over the values of $\Delta$, we can map out the space of its allowed values (see Appendix \ref{app:spectralgap} for further details). In practice, we observe that the bootstrap-allowed regions for $\Delta$ form a single interval, thus providing lower and upper bounds on $\Delta$.

\begin{figure}[h!]
    \centering
    \includegraphics[width=0.80\linewidth]{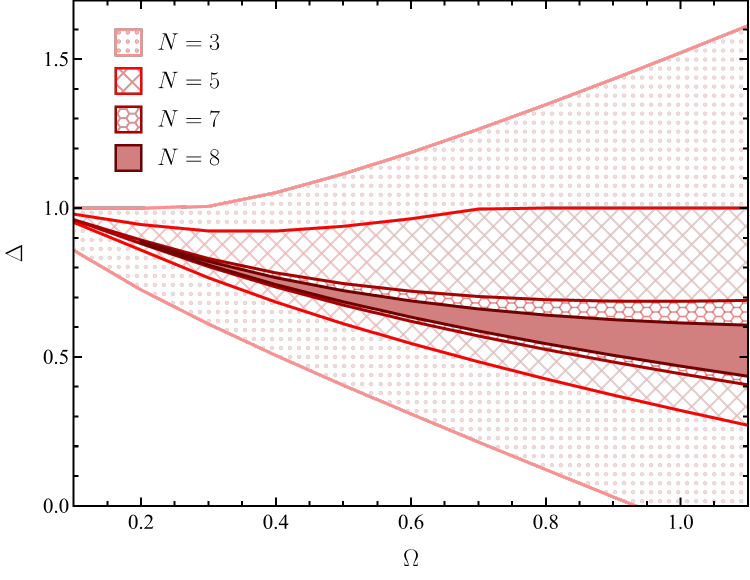}
    \caption{Bootstrap-allowed regions (shaded) for the spectral gap $\Delta$ as functions of $\Omega$.}
    \label{fig:pltGap}
\end{figure}

\begin{figure}[h!]
    \centering
    \includegraphics[width=0.32\linewidth]{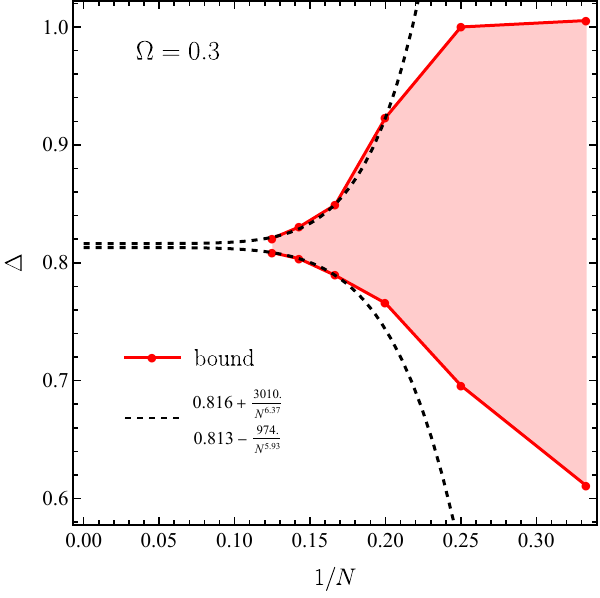}
    \includegraphics[width=0.32\linewidth]{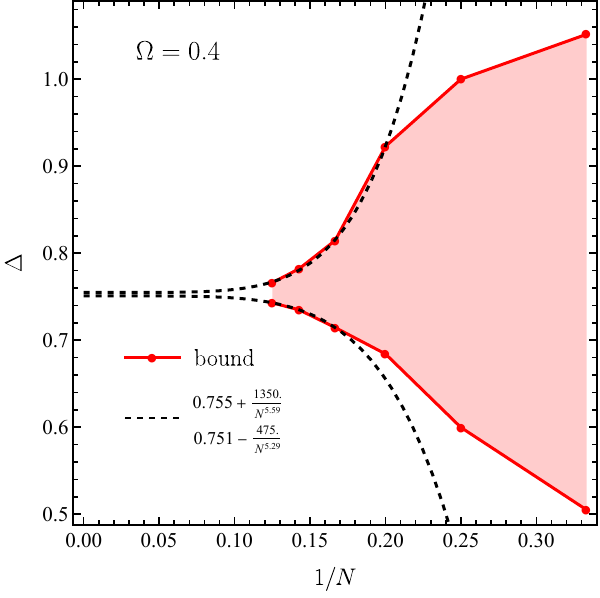}
    \includegraphics[width=0.32\linewidth]{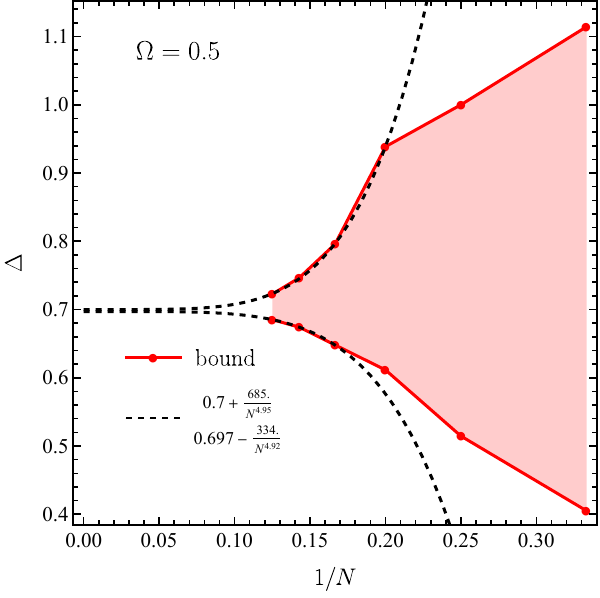}
    \caption{We show the how bootstrap bounds on $\Delta$ shrink as $N$ increases, for $\Omega=0.3$, $\Omega=0.4$ and $\Omega=0.5$ as representative examples.  }
    \label{fig:pltGap_2}
\end{figure}

In fig.~\ref{fig:pltGap}, we show the bootstrap bounds on $\Delta$ for several values of $\Omega$ for different subsystem sizes $N$. Not surprisingly, the bounds become tighter as $\Omega$ moves towards $\Omega=0$. In fig.~\ref{fig:pltGap_2} we further present how bootstrap bounds on $\Delta$ depend on the inverse subsystems size $1/N$ for several representative values of $\Omega$. Using three or four points with largest $N$, we extrapolate the gap dependence on $1/N$ based on simple polynomial functions. This produces rough and non-rigorous estimates for the values of the gap $\Delta$ in the $N\to \infty$ limit.

In table~\ref{tab:gap}, we show bootstrap bounds on the gap at $\Omega = 2.00$ for several subsystem sizes, together with the estimated gap from the data found in \cite{Carollo:2019zmt,Gillman:2019lfe} using the matrix product states and the time-evolving block decimation algorithm.\footnote{We greatly appreciate Federico Carollo for providing us with the data presented in \cite{Carollo:2019zmt,Gillman:2019lfe}.} Clearly, our bounds are consistent with the prediction in \cite{Carollo:2019zmt,Gillman:2019lfe}, but in order to make the bounds tighter, one needs to impose larger $N$ bootstrap constraints.

\begin{table}[h!]
\centering
\begin{tabular}{lcccc}
& $N=6$ & $N=7$ & $N=8$ & estimate from \cite{Carollo:2019zmt,Gillman:2019lfe} \\ \hline
Lower bound & 0.032336 & 0.127059 & 0.186932 & \multirow{2}{*}{$\approx 0.316$} \\
Upper bound & 1.000000 & 0.791741 & 0.629361 & \\ \hline
\end{tabular}
\caption{Bounds on the spectral gap computed at $\Omega = 2.0$ with different subsystem sizes, compared with the gap extracted from \cite{Carollo:2019zmt,Gillman:2019lfe}.}
\label{tab:gap}
\end{table}

\section{Outlook}
In this work, we established a bootstrap framework for accessing the physics of open quantum many-body systems with absorbing phase transitions. There are several clear directions for further investigations.

First, although the bootstrap bounds obtained in this work are rigorous, they have not yet converged to the estimates obtained by other methods, such as those in \cite{Gillman:2019lfe,Jo:2021tax}. The main practical obstacle is the exponentially growing number of bootstrap variables and constraints as the size of the sublattice on which the bootstrap is imposed increases. In \cite{Cho:2024owx}, a coarse-grained bootstrap approach was developed for closed quantum systems, in which only a linearly growing but most relevant subset of bootstrap variables and constraints is selected and imposed. However, when we applied its natural extension to Lindblad dynamics, particularly in the quantum contact process, we did not observe any visible improvement in the bootstrap bounds compared with the results presented in this work. The key question, then, is which subset of bootstrap constraints is most relevant for nonequilibrium quantum systems. Unlike low-energy states in one-dimensional spin chains, which admit faithful tensor-network representations, nontrivial steady states of Lindblad dynamics generally need not allow such faithful coarse-grained descriptions, and may even exhibit genuine volume-law entanglement entropy.

Second, the bootstrap framework developed in this work can be extended straightforwardly to open quantum many-body systems with discrete time evolution, such as quantum circuits. In particular, quantum circuits with absorbing states, such as the one discussed in \cite{ODea:2022wzh}, are natural candidates for the bootstrap approach. Indeed, classical circuits such as probabilistic cellular automata were recently studied using bootstrap methods in \cite{Cho:2025dgc}, and that strategy can be naturally combined with the framework developed here to bootstrap quantum circuits.

Finally, it would be desirable to develop a bootstrap method that can directly access information-theoretic quantities in open quantum many-body systems, rather than only expectation values of local operators. Given the recent discovery of various nonequilibrium phase transitions of an information-theoretic nature, the development of such a method would be especially timely. Recent bootstrap approaches to thermal equilibrium \cite{Fawzi:2023fpg,Cho:2024kxn,Cho:2025vws} already access the free energy, which contains the entropy contribution, suggesting that other information-theoretic quantities may also be accessible within the bootstrap framework.

\section*{Acknowledgements}
We are very grateful to Federico Carollo for sharing with us the data presented in \cite{Carollo:2019zmt,Gillman:2019lfe}. The work of M.C. is supported by Clay C\'ordova's Sloan Research Fellowship from the Sloan Foundation. The work of P.T. has been supported in parts through a joint UK Research and Innovation (UKRI) partnership with the US National Science Foundation (NSF) under grant no.~EP/Z003423/1.

\appendix

\section{Numerical details in spectral gap search}\label{app:spectralgap}

In the implementation of the spectral gap search, the way to rewrite the feasibility problem as an optimization problem has been discussed in the literature as the navigator function method~\cite{Reehorst:2021ykw}, where the navigator function is defined as the minimal value of $g$ in~\eqref{gap-search} 
\begin{equation}
    \mathcal{N}(\Delta) \equiv g_{\rm min}(\Delta).
\end{equation}
A positive/negative navigator function rules out/in the gap value. The root searching algorithm has two phases. In phase one it looks for $\underset{\Delta}{\rm argmin}\,\mathcal{N}(\Delta)$ and exits upon finding a negative $\mathcal{N}$. In phase two, it uses Brent method to search for the boundary of the allowed region $\mathcal{N}(\Delta) = 0$ on both sides.
\begin{figure}[h!]
    \centering
    $\begin{array}{cc}
    \raisebox{-0.8in}{\includegraphics[width=0.49\linewidth]{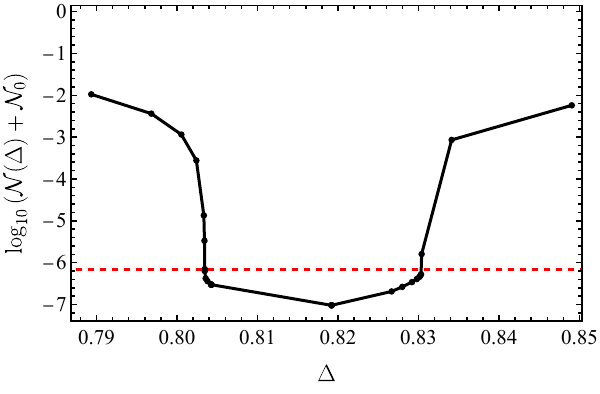}}~~
    &
    \includegraphics[width=0.3\linewidth]{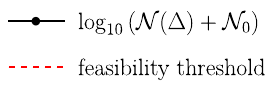}
    \end{array}$
    \caption{\label{sep-scales}Plot of the shifted navigator function $\mathcal{N}(\Delta)+\mathcal{N}_0$ in log scale at $\Omega = 0.3$ and $L = 7$. The shift is $\mathcal{N}_0= 7.0\times 10^{-7}$. By definition of the navigator function, $\mathcal{N}(\Delta)<0$ means $\Delta$ is allowed, and $\mathcal{N}(\Delta)>0$ means $\Delta$ is disallowed. The zero point of $\mathcal{\Delta}$ is represented by the red dashed line. $\Delta$ is only allowed at a small band around $0.82$, and inside the allowed region the navigator function takes a tiny negative value of $\mathcal{N}(\Delta) \approx -10^{-6}$, while the disallowed region generally takes $\mathcal{N}(\Delta)\sim 10^{-2}$. }
    \label{fig:feasibility}
\end{figure}

We notice the navigator function has a wide separation of scales. An example is shown in fig \ref{sep-scales}, where at $\Omega = 0.3$ and $L=7$ the gap is only allowed for a narrow band in $\Delta$ and the allowed region has an absolute navigator value $10^5$ times smaller than the disallowed region. The separation of scale gets worse for larger $L$ and smaller values of $\Omega$. This means, for a semidefinite solver with machine precision, the negative value signal may be below the numerical noise, and the Newton's method for finding the boundary of the allowed region may also be influenced by the finite precision, causing a false bound. For $\Omega=0.1$ we encounter this issue at $L=8$ but the other parameters are fine at the levels that we study.

\FloatBarrier
\bibliographystyle{JHEP}
\bibliography{sample}

\end{document}